\documentclass[aps, reprint, amsmath, amssymb, amsfonts, floatfix, intlimits,superscriptaddress]{revtex4-1}
\pdfoutput=1
\usepackage{graphicx}
\usepackage{dcolumn}
\usepackage{bm}
\usepackage{color} 
\usepackage{float}

\definecolor{deepBlue}{RGB}{48, 71, 210}
\definecolor{myGreen}{RGB}{200, 38, 216}
\definecolor{APSBlue}{RGB}{46, 48, 146}
\usepackage{url}
\usepackage{color}
\usepackage[colorlinks = true,
	linkcolor = blue,
	urlcolor  = blue,
	citecolor = blue,
	anchorcolor = blue]{hyperref}

\begin{document}
	
	\title{Moir\'e-less Correlations in ABCA Graphene}
	
\affiliation{
	Department of Physics, Columbia University, New York, New York 10027, United States\looseness=-1}
\affiliation{
	Max Planck Institute for the Structure and Dynamics of Matter, Luruper Chaussee 149, 22761 Hamburg, Germany\looseness=-1}
\affiliation{
	Institut fur Theorie der Statistischen Physik, RWTH Aachen University, 52056 Aachen, Germany and JARA-Fundamentals of Future Information Technology, 52056 Aachen, Germany\looseness=-2}
\affiliation{
	Department of Mechanical Engineering, Columbia University, New York, NY, USA\looseness=-1}
\affiliation{
	National Institute for Materials Science, 1-1 Namiki, Tsukuba 305-0044, Japan\looseness=-1}
\affiliation{
	Center for Computational Quantum Physics (CCQ), The Flatiron Institute, 162 Fifth Avenue, New York, NY 10010, USA\looseness=-1}
\affiliation{
	Nano-Bio Spectroscopy Group, Departamento de Fisica de Materiales, Universidad del País Vasco, 20018 San Sebastian, Spain\looseness=-1}

\author{Alexander Kerelsky}
\altaffiliation{These authors contributed equally to this work.
}
\affiliation{
	Department of Physics, Columbia University, New York, New York 10027, United States\looseness=-1}
\author{Carmen Rubio-Verd\'u}
\altaffiliation{These authors contributed equally to this work.
}
\affiliation{
	Department of Physics, Columbia University, New York, New York 10027, United States\looseness=-1}
\author{Lede Xian}
\affiliation{
	Max Planck Institute for the Structure and Dynamics of Matter, Luruper Chaussee 149, 22761 Hamburg, Germany\looseness=-1}
\author{Dante M. Kennes}
\affiliation{
	Institut fur Theorie der Statistischen Physik, RWTH Aachen University, 52056 Aachen, Germany and JARA-Fundamentals of Future Information Technology, 52056 Aachen, Germany\looseness=-2}
\author{Dorri Halbertal}
\affiliation{
	Department of Physics, Columbia University, New York, New York 10027, United States\looseness=-1}
\author{Nathan Finney}
\affiliation{
	Department of Mechanical Engineering, Columbia University, New York, NY, 		USA\looseness=-1}
\author{Larry Song}
\affiliation{
	Department of Physics, Columbia University, New York, New York 10027, United States\looseness=-1}
\author{Simon Turkel}
\affiliation{
	Department of Physics, Columbia University, New York, New York 10027, United States\looseness=-1}
\author{Lei Wang}
\affiliation{
	Department of Physics, Columbia University, New York, New York 10027, United States\looseness=-1}
\author{K. Watanabe}
\affiliation{
	National Institute for Materials Science, 1-1 Namiki, Tsukuba 305-0044, Japan\looseness=-1}
\author{T. Taniguchi}
\affiliation{
	National Institute for Materials Science, 1-1 Namiki, Tsukuba 305-0044, Japan\looseness=-1}
\author{James Hone}
\affiliation{
	Department of Mechanical Engineering, Columbia University, New York, NY, USA\looseness=-1}
\author{Cory Dean}
\affiliation{
	Department of Physics, Columbia University, New York, New York 10027, United States\looseness=-1}
\author{Dmitri Basov}
\affiliation{
	Department of Physics, Columbia University, New York, New York 10027, United States\looseness=-1}
\author{Angel Rubio}
\altaffiliation{Correspondence to:
	\href{mailto:apn2108@columbia.edu}{apn2108@columbia.edu} (A.N.P); \\
	\href{mailto:angel.rubio@mpsd.mpg.de}{angel.rubio@mpsd.mpg.de} (A.R.)}
\affiliation{
	Max Planck Institute for the Structure and Dynamics of Matter, Luruper Chaussee 149, 22761 Hamburg, Germany\looseness=-1}
\affiliation{
	Center for Computational Quantum Physics (CCQ), The Flatiron Institute, 162 Fifth Avenue, New York, NY 10010, USA\looseness=-1}
\affiliation{
	Nano-Bio Spectroscopy Group, Departamento de Fisica de Materiales, Universidad del País Vasco, 20018 San Sebastian, Spain\looseness=-1}
\author{Abhay N. Pasupathy}
\altaffiliation{Correspondence to:
	\href{mailto:apn2108@columbia.edu}{apn2108@columbia.edu} (A.N.P); \\
	\href{mailto:angel.rubio@mpsd.mpg.de}{angel.rubio@mpsd.mpg.de} (A.R.)}
\affiliation{
	Department of Physics, Columbia University, New York, New York 10027, United States\looseness=-1}

	\date{\today}
	\begin{abstract}
		Atomically thin van der Waals materials stacked with an interlayer twist have proven to be an excellent platform towards achieving gate-tunable correlated phenomena linked to the formation of flat electronic bands. In this work we demonstrate the formation of emergent correlated phases in multilayer rhombohedral graphene - a simple material that also exhibits a flat electronic band but without the need of having a moir\'e superlattice induced by twisted van der Waals layers. We show that two layers of bilayer graphene that are twisted by an arbitrary tiny angle host large (micron-scale) regions of uniform rhombohedral four-layer (ABCA) graphene that can be independently studied. Scanning tunneling spectroscopy reveals that ABCA graphene hosts an unprecedentedly sharp flat band of 3-5 meV half-width. We demonstrate that when this flat band straddles the Fermi level, a correlated many-body gap emerges with peak-to-peak value of 9.5 meV at charge neutrality. Mean field theoretical calculations indicate that the two primary candidates for the appearance of this broken symmetry state are a charge transfer excitonic insulator and a ferrimagnet. Finally, we show that ABCA graphene hosts surface topological helical edge states at natural interfaces with ABAB graphene which can be turned on and off with gate voltage, implying that small angle twisted double bilayer graphene is an ideal programmable topological quantum material.
	\end{abstract}
	\maketitle
	
	Two dimensional (2D) van der waals (vdW) heterostructures with an interlayer twist have provided a new avenue for observing emergent tunable many-body electron phenomena. Recent experimental realizations include twisted bilayer graphene (tBG) near the so-called “magic angle” of 1.1$^\circ$ \cite{cao18i,cao18s,yankowitz19}, twisted double bilayer graphene (tDBG) at a magic angle of around 1.3$^\circ$ \cite{shen19,liu19,cao19}, ABC trilayer graphene on near-perfectly aligned hexagonal boron nitride (hBN) (ABC-tLG/hBN) \cite{chen19i,chen19s} and transition metal dichalcogenide heterostructures \cite{tang19,regan19,an19,wang19} (with predictions on a variety of other systems \cite{xian19,kennes19}). All of these systems host an interplay of two phenomena - the presence of one or more van Hove singularities (which we colloquially refer to as “flat bands” henceforth) at low energy where the density of states is sharply peaked, and the existence of a moir\'e pattern that creates a unit cell that is about a hundred times larger than the carbon - carbon nearest neighbor distance in graphene. The large number of electrons with quenched kinetic energy make the flat bands conducive to interaction driven phases \cite{bist11}. The enlarged moir\'e unit cell is thought to reduce both the flat band bandwidth and the interaction energy scales, and also introduces easily accessible integer fillings that create Mott-like insulating states \cite{cao18i,cao18s,yankowitz19,shen19,liu19,cao19,chen19i,chen19s,tang19,regan19,an19,wang19}, the relation of which to nearby superconductivity is debated. A natural question that arises from all of these works is whether the moir\'e pattern is a necessary condition for the observation of correlated many-body phases, or whether it is simply sufficient to further reduce the flat band bandwidth and hence the kinetic energy in the heterostructure.

	\begin{figure*}[t]
		\includegraphics[width=\linewidth]
		{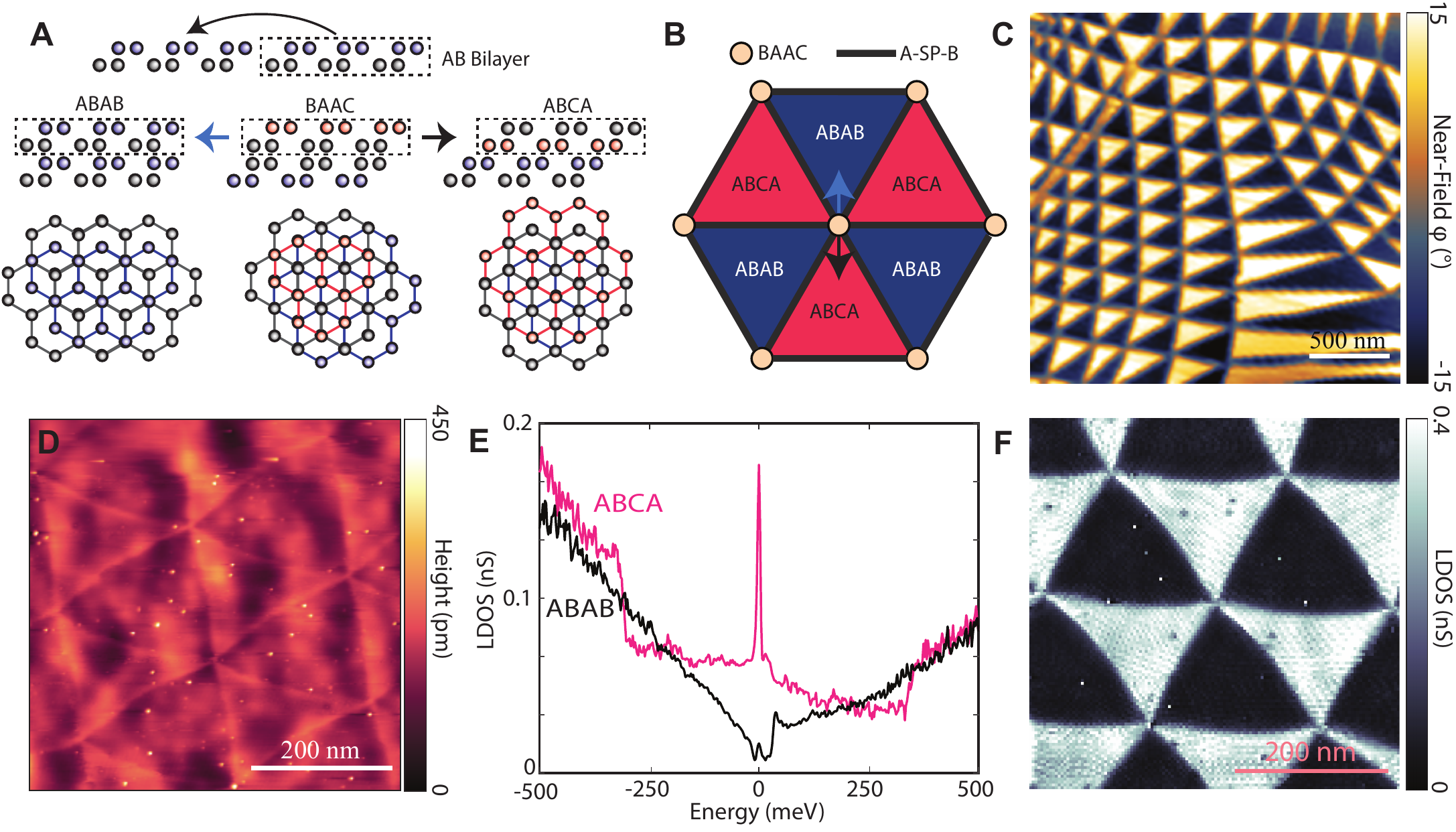}
		\caption{
			\textbf{Rhombohedral Domains in Small Angle Twisted Double Bilayer Graphene.} (\textbf{A}) Side view and top view of atomic stacking considerations in a tDBG moir\'e pattern. Due to the enforced Bernal stacking of the top and bottom layer, the AA site stacking equivalent in tDBG is BAAC. One can translate the bilayer - bilayer atomic alignment by one atom in either direction to visualize commensurate stackings around the BAAC site (as shown by arrows in Figure 1B for a moir\'e unit cell diagram). While the monolayer - monolayer case (which can be seen by the middle two layers) becomes AB/BA, when the top and bottom tDBG layers are considered the commensurate stacking sites become ABAB (Bernal) and ABCA (rhombohedral). (\textbf{B}) Cartoon moir\'e of small angle tDBG with two example directions of atomic stacking shift considered in (A). (\textbf{C}) A nano-infrared image of a large area of a small-angle tDBG sample. (\textbf{D}) An STM topographic image of small angle tDBG (setpoints of 300 mV and 100 pA). (\textbf{E}) STS LDOS of ABCA and ABAB graphene domains in small angle tDBG taken at the center of one of each domain (setpoints of 300 mV and 150 pA with a lock-in oscillation of 2.5 mV). (\textbf{F}) STS LDOS map at the energy of the ABCA flat band (setpoints of 400 mV and 150 pA with a 2.5 mV oscillation).} 
	\end{figure*} 
	
	In this regard, multilayer rhombohedral (ABC) graphene offers a different perspective towards achieving flat bands without the use of a moir\'e potential \cite{zhang10}. Indeed, in a seminal work \cite{min08}, it was theoretically shown that the low energy band structure of multilayer rhombohedral graphene has a sharply peaked DOS, with the band structure $E(k)\propto k^N $ (where N is the number of layers) at low energy in the nearest neighbor hopping approximation. This implies a peak in the DOS at charge neutrality in this material for N$>$2, with an appreciable fraction of the entire band within this peak \cite{yelgel16}. Indeed, this physics is already at play in ABC-tLG/hBN \cite{chen19i,chen19s}, where some of the flatness of the bands comes from the intrinsic band structure of ABC graphene, which is then further flattened and isolated by the moir\'e pattern from the hBN alignment. A facile alternative to flatten the bandwidth without introducing a moir\'e potential is to simply increase the number of layers of the rhombohedral stacked graphene. Unfortunately, isolating rhombohedral stacked graphene of any thickness is extremely difficult as it is less energetically favorable than the multilayer counterpart, Bernal stacked graphene. Since the difference between rhombohedral and Bernal graphene is simply a lattice shift, and the interlayer van der Waals forces are weak, it is well known that rhombohedral graphene reverts to the Bernal form when samples are processed with heat, pressure or while performing lithography \cite{sugawara18}. In this work, we show that twisting two sheets of bilayer graphene (tDBG) by a tiny ($<$0.1$^\circ$) angle is a simple and robust method to create large area (up to micron-scale) rhombohedral graphene of four-layer thickness (ABCA graphene). We present gate tunable scanning tunneling microscopy and spectroscopy (STM/STS) measurements at 5.7 Kelvin on these regions. We show that correlated phases can be achieved without the need for a moir\'e superlattice potential and that rhombohedral graphene has unique topological properties. 
	
	In the limit of small twist angles, it is well known that two twisted layers of graphene will prefer to form extended regions where the lattice is commensurate and Bernal stacked (called AB or BA stacking) \cite{yoo19}, separated by sharp domain walls where the stacking order changes from AB to BA \cite{wong15,huang18,morgen}. Where the domain walls intersect, the stacking order is locally AA, which is also where the low-energy electrons are localized \cite{bist11,kerelsky19}. We begin by considering the atomic stacking nature of small angle tDBG, as shown in Figure 1A-B. The easiest way to think of tDBG is to consider the middle two layers which form the same structure as tBG. The rest of the structure of tDBG can then be visualized by the fact that the two upper and two lower layers are both Bernal stacked. Using this visualization technique, it becomes clear that the equivalent of the AA stacking location in tBG is a BAAC stacking in tDBG. One can then displace the top bilayer by one atom in either direction to visualize the other commensurate stacking sites in a manner exactly analogous to tBG. The consequence of this as shown in Figure 1A is the formation of ABAB (Bernal) and ABCA (rhombohedral) stacked four-layer graphene in the commensurate regions of the sample. Because it is energetically favorable for commensurate regions to maximize in area, small angle tDBG has large, uniform extended regions of rhombohedral and Bernal graphene that push any effect of the tDBG moir\'e superlattice to the interfaces of the commensurate domains. At small angles, tDBG thus forms large, uniform rhombohedral domains that are decoupled from effects of the large wavelength small angle tDBG moir\'e. The final stacking arrangement that arises is then shown in Figure 1B. The rhombohedral graphene is protected by the global symmetry enforced by the twist angle, making it stable and easy to produce. While in this work we show this case for small angle tDBG, the atomic stacking arguments at hand will also extend to any angle including the magic angle of tDBG where the ABCA sites will likely play a role in electron localization along with the BAAC sites. These arguments also apply to monolayer-bilayer twisted trilayer graphene where the commensurate stackings become ABA and ABC \cite{morgen} ($(60+\delta)^\circ$ tDBG is another interesting case where commensurate sites will be ABCB and BCBA). 
	
	We first use nano-infrared imaging to visualize small angle tDBG on a large scale. Figure 1C shows a nano-infrared image of a large area region of small angle tDBG (details in supplementary materials). We see that there are two distinct sets of domains (bright and dark in the image) with quasi-hexagonal order. The average size of the domains in this sample is of order 200 nm on a side, though several larger domains of over a micron in size are seen, presumably due to local strain or pinning of particular stacking sites. We have observed such domain structures of hundreds of nanometers in width in every small angle tDBG sample. The contrast between the two domains clearly shows that the two domains have distinct electronic responses, as we might expect for Bernal and rhombohedral graphene. Having seen the optical signatures of small angle tDBG structure on a large length scale, we turn to STM/STS to study the atomic-scale spectroscopic properties. In Figure 1D we show a STM topographic image of a homogenous small angle tDBG region with triangular domains that are about 220 nm per side. This can directly be contrasted with a small angle tBG STM topography in Figure S1 (supplementary materials). In tBG, the STM topographic contrast is dominated by the AB/BA domain walls and the AA site centers, while the two commensurate AB and BA domains have identical contrast. In the case of tDBG, the contrast differs between the two sets of triangular domains, ABCA and ABAB. 
	
	We next turn to the electronic structure of the domains. In Figure 1E we show STS LDOS on the ABCA and ABAB domains elucidating their starkly different electronic properties. While the ABAB domains show a gradually increasing density of states away from the Fermi level (consistent with measurements on multilayer graphite), the ABCA domains have an extremely sharp enhancement in the density of states near the Fermi level. In Figure 1F we show an STS LDOS image at the energy of the LDOS peak. This image shows that the contrast between the ABCA and ABAB domains is uniformly observed in the low energy electronic structure, and thus the peak in the LDOS in Figure 1E is simply a property of the ABCA graphene itself and not due to confinement or impurity effects. Furthermore, the uniform spatial electronic structure proves that these ABCA domains have spatially maximized their commensurate area, pushing the unnatural atomic stackings of the spatially evolving tDBG moir\'e to the domain walls and decoupling the ABCA graphene from the effects of the superlattice potential. Thus, although we use a tDBG moir\'e to achieve the ABCA graphene regions, they are no longer affected by the much larger moir\'e, allowing the study of ABCA graphene without any superlattice potential affecting its low-energy electronic properties. From the topography and STS map in Figs. 1D and F, it can also be seen that the ABAB Bernal domains are convex whereas the ABCA domains are concave. This domain shape can simply be understood as a competition between the desire on the part of the ABAB domains to maximize their area (being more energetically favorable) and the desire to minimize the length of the domain wall (similar to surface tension). This behavior is also different from tBG, where the two AB/BA domains are equivalent, leading to simple straight domain walls as seen in Figure S1 and previous works \cite{huang18,kerelsky19}. 
	
	\begin{figure*}[t]
		\includegraphics[width=\linewidth]
		{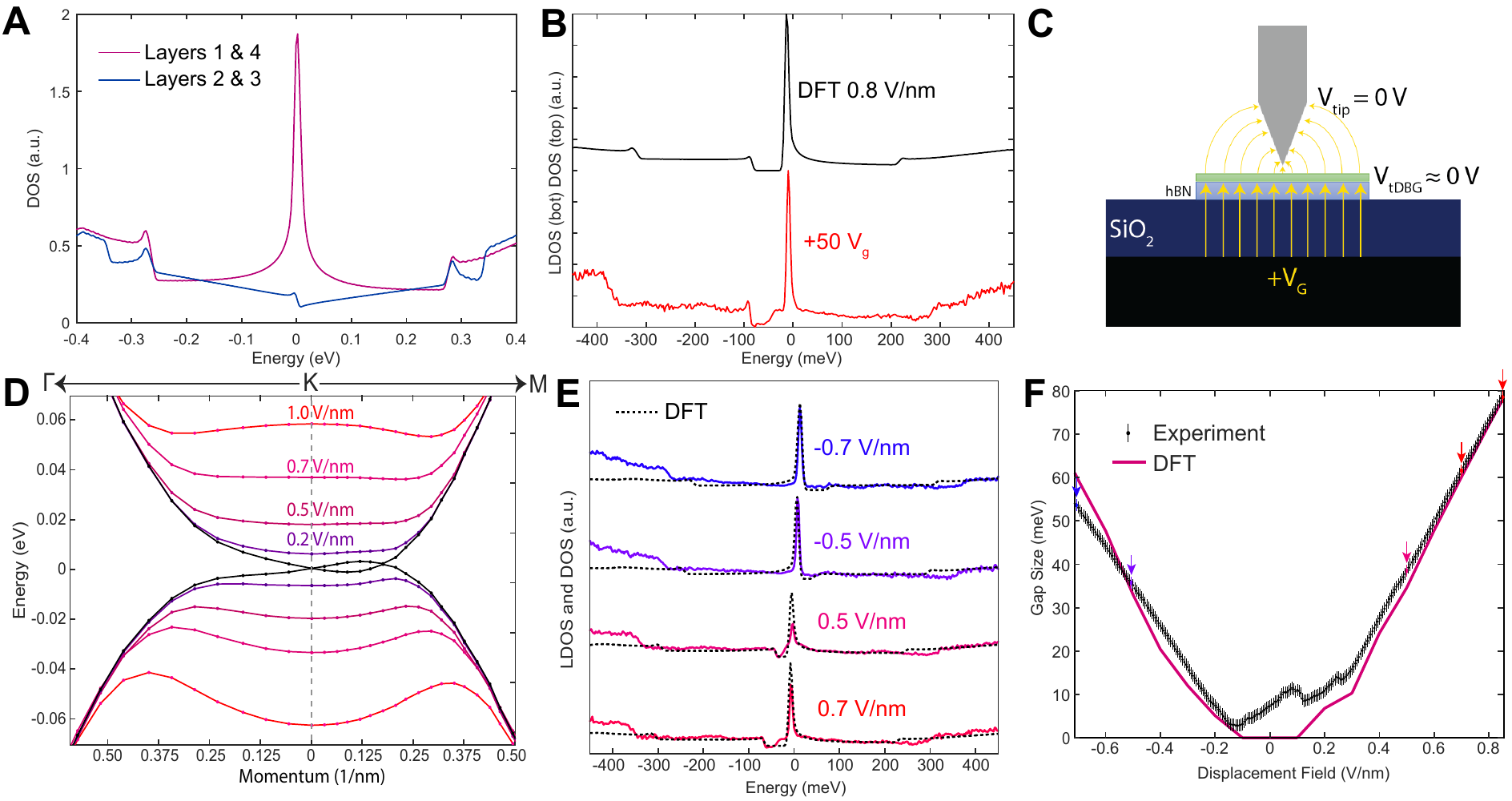}
		\caption{
			\textbf{Displacement Field Dependent LDOS in ABCA Graphene.} (\textbf{A}) Layer-dependent LDOS as calculated by self-consistent DFT. (\textbf{B}) STS LDOS curve at a gate voltage of 50 V and top layer DFT calculated DOS at 0.8 V/nm. (\textbf{C}) Cartoon of the STM/STS measurement geometry under an applied back gate voltage. The sample is biased relative to the grounded tip. The back gate dopes the sample and inherently also induces a displacement field. (\textbf{D}) Displacement field dependent low-energy band structure of ABCA graphene as calculated by DFT. (\textbf{E}) A series of experimental STS LDOS (solid) and DFT top layer LDOS (dotted) curves at various displacement field magnitudes. (\textbf{F}) Extracted displacement field gap sizes for 251 STS LDOS curves equally spaced between our displacement field extremes. DFT calculated displacement field gap sizes are overlaid. STS measurements in (B) and (E) were at setpoints of 400 mV and 150 pA with a lock-in oscillation of 2.5 mV. } 
	\end{figure*}

	Having clearly identified the large regions of uniform ABCA graphene in our sample, we now focus on the low energy gate-dependent spectroscopic structure. It is important to note that in STM measurements, the spectroscopic signals are dominated by the first conducting layer of the sample, and hence by the top layer of ABCA graphene. To begin, we compare the surface measured experimental LDOS (Figure 1E) to layer-dependent theoretical LDOS as calculated by DFT and shown in Figure 2A. The low energy density of states is concentrated on the outer layers (i.e., layers 1 \& 4) and we find that when examined over a large energy range, the LDOS predicted by DFT for the top layer matches the observed spectroscopy in Figure 1E strikingly well, including the high energy bands appearing as steps in the spectrum. 
	
	We now turn to the effect of a gate voltage on our measurements. In Figure 2B we show a curve at our maximum positive gate voltage of 50 V (limited by gate leakage). We see that the shape of the measured spectrum has undergone significant changes. The low energy peak and high energy band edges are still present in the spectrum (though they have shifted due to doping), but in addition a clear gap has emerged below the Fermi level. To better understand the origin of this gap, in Figure 2C we show a model for our geometry and basic electrostatics. In our measurement, the sample is biased relative to the tip, while the gate voltage is applied relative to the sample. The application of a back gate voltage results in both electrostatic doping of the sample as a whole, as well as an electric field applied across the layers of the sample. In trilayer ABC layer graphene \cite{zou13}, it has indeed been shown that electric fields can cause a gap to form. To better understand this effect in four-layer ABCA graphene, we calculate the electric field dependence of the band structure by means of self-consistent DFT (see supplementary materials) in Figure 2D. Upon application of a displacement field to the sample, a gap emerges in the ABCA graphene band structure as seen in Figure 2B, which shows the DFT calculated top layer DOS on top. While the band structure of the top layer is electron-hole asymmetric in an applied displacement field, the bottom layer has a band structure that is approximately electron-hole mirror symmetric with respect to the top layer. We also note that in the presence of a displacement field, the low-energy bands develop a topological character, with each of the valleys in the conduction and valence bands having a Chern number of 2, but with opposing signs \cite{zhang13,li10} (see Figure S2 and supplementary materials). 
			
	\begin{figure*}[t]
		\includegraphics[width=\linewidth]
		{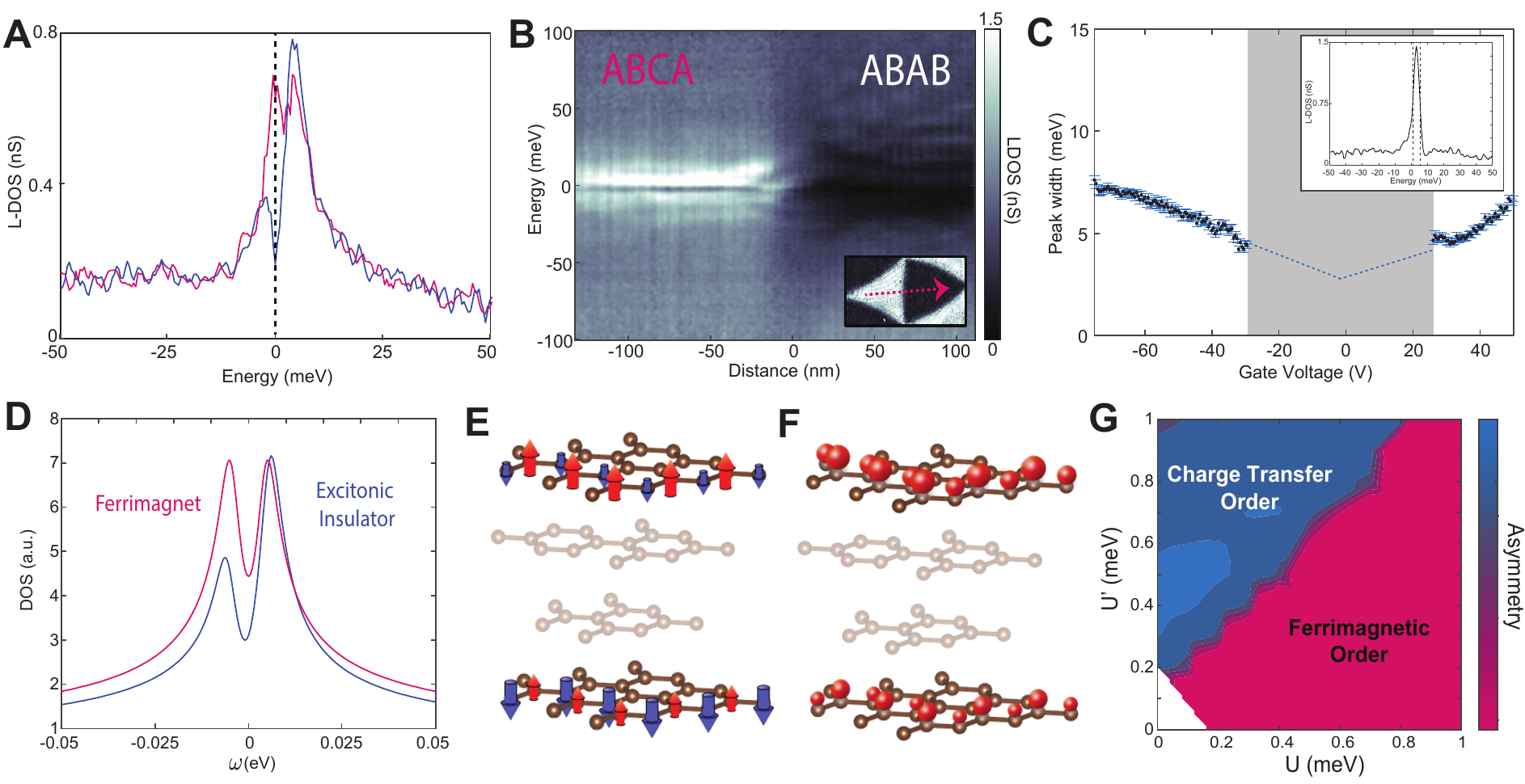}
		\caption{
			\textbf{Electronic Correlations in ABCA Graphene.} (\textbf{A}) Two STS LDOS curves at small displacement fields (blue at neutrality and approximately 0 V/nm and magenta at -0.1 V/nm) where in a single-particle picture there should be no gap. The flat band splits and shows a correlated gap of 9.5 meV and 4.5 meV respectively. (\textbf{B}) STS LDOS linecut at charge neutrality across an ABCA and an ABAB domain. (\textbf{C}) ABCA flat-band half width as a function of gate voltage. The shaded region in grey is excluded as the flat band half-width cannot be cleanly extracted due to splitting. The inset shows an STS spectrum with the peak still far from the Fermi level, exhibiting a 4 meV half-width. STS measurements in (A-C) were taken at setpoints of 300 mV and 200 pA with a lock-in oscillation of 0.5 mV. (\textbf{D}) Tight binding mean field DOS for ferrimagnetic and charge transfer excitonic insulator ordered states. (\textbf{E}-\textbf{F}) Visual representation of the two ordered states in ABCA graphene. The mostly irrelevant middle layers are shaded. Ferrimagnetic spin order (E) and charge transfer insulating order (F) are shown by arrows pointing up or down as well as big or small spheres, respectively (shown on select sites only). (\textbf{G}) Asymmetry of peaks in mean-field tight binding model as a function of $U$, intralayer, and $U'$, top-bottom layer interactions. The most likely ordered state in each region is labeled. } 
	\end{figure*} 

	For the STS curve at +50 V$_G$ in Figure 2B, we estimate our displacement field to be 0.85 V/nm (we assume the tip to be one electrode of a parallel plate capacitor held at 0 V and account for intrinsic sample doping which displaces charge neutrality and the zero field condition). The DFT prediction for the top layer DOS with an electric field of 0.8 V/nm (solid black curve on Figure 2B) matches the experimental spectrum well. We note that in the DFT calculation, the electric field value is fixed at infinity, and the charge on the layers is allowed to redistribute itself self-consistently.  We have measured the experimental LDOS for a range of gate voltages, a few of which are displayed in Figure 2E. Overlaid on these curves are the DFT predictions at the corresponding electric field values. As the field is increased, the displacement field gap grows in size. For electric fields pointing from the gate to the tip, the gap is located to the left (lower energy) of the main peak in the LDOS, while for fields pointing in the other direction the gap is located to the right (higher energy). In Figure 2F we track the displacement field dependent gap size for 251 STS LDOS curves extending between our gate voltage extremes of -75 V and 50 V with DFT calculated gap size overlaid. For displacement field magnitudes larger than about 0.3 V/nm, DFT and experiment are in excellent agreement. The curves shown in Figure 2E are all in this category. While the match between experiment and DFT is excellent at high fields in Figure 2F, for low fields the single particle DFT fails to capture the experimental results. In particular, DFT predicts that for electric fields between -0.1 to 0.1 V/nm, the gap produced by the displacement field is negligible. This is a consequence of screening that reduces the field inside the ABCA graphene. The experimental data is in stark contrast to this prediction, always displaying a non-zero gap in this region that is of order 5-10 meV. The fact that the gap matches quantitatively at high positive and negative field negates any possibility of extraneous fields (for example from the tip) affecting this conclusion. We are thus forced to consider other explanations for the observed gap at charge neutrality in experiment. 
	
	To highlight the gap at zero field, in Figure 3A, we show two LDOS curves at small displacement field values. The curve shown in blue is exactly at charge neutrality (and 0 field) with a gap of 9.5 meV centered precisely at the Fermi level. The curve shown in magenta is the smallest gap observed at any value of gate voltage, 4.5 meV in magnitude, measured at a field of about -0.1 V/nm. This gap in the LDOS is consistent throughout the entire ABCA domain as can be seen in Figure 3B where we plot the LDOS at a sequence of points across a whole ABCA and ABAB domain at charge neutrality. Given the clear disagreement with single particle DFT calculations, this gap is a consequence of electron correlations. One question that arises is why ABCA graphene displays such a correlated gap, while ABC graphene has not displayed this phenomenon in previous measurements \cite{yin19}. To answer this question, we quantify the sharpness of the ABCA LDOS peak by extracting the half-width of the most prominent peak in the spectrum at each value of gate voltage. A plot of these widths as a function of gate voltage is shown in Figure 3C. For a range of gate voltages near charge neutrality (grey region in Figure 3C), the presence of the correlated gap in the peak makes it difficult to define a peak width cleanly. At larger gate voltages, we find that the peak width increases linearly with gate voltage, likely a quasiparticle lifetime effect. An extrapolation of the linear behavior to charge neutrality shows that without splitting, the peak would have a half width of 3 meV when the sample is charge neutral. Previous measurements of ABC graphene have shown a peak width of 10-20 meV near charge neutrality \cite{min08} and about 50 meV on a conducting substrate \cite{pierucci16}. This sharpness of the peak in the DOS of ABCA graphene is far lower than trilayer ABC graphene (as expected in the low energy approximation) or even the flat bands of magic angle tBG \cite{kerelsky19,choi19,xie19,jiang19}, which are about 10 meV in half width \cite{kerelsky19}. It is important to reiterate that at these small angles in tDBG, the commensurate ABCA regions are energetically maximized and have pushed any effects of the tDBG moir\'e to the domain walls, thus this finding is purely associated with ABCA graphene. ABCA graphene is thus a simple flat band system where kinetic energy is sufficiently quenched such that a correlation induced gap emerges without a relevant moir\'e periodicity. 
		
	\begin{figure*}[t]
		\includegraphics[width=\linewidth]
		{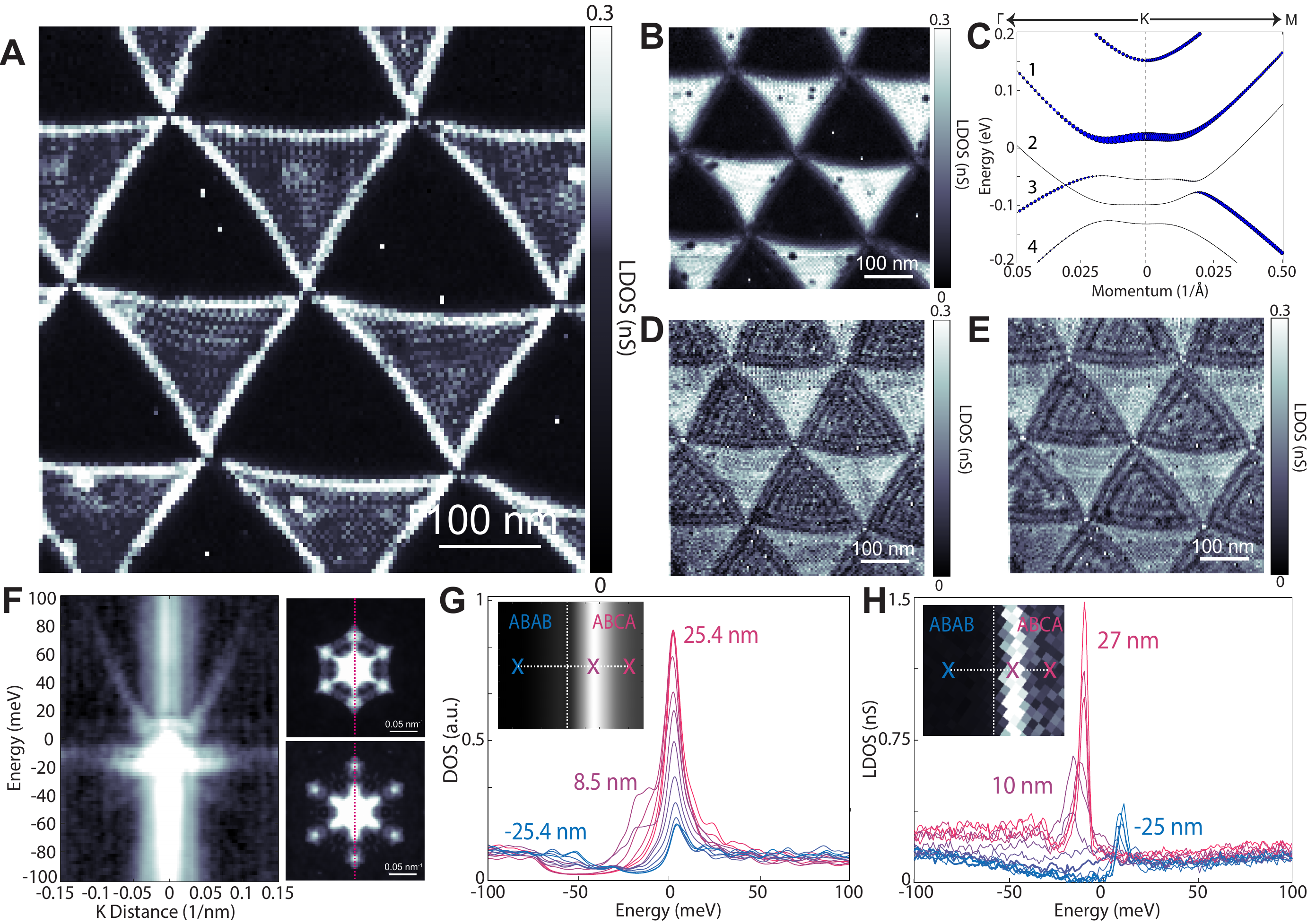}
		\caption{
			\textbf{Gate-tunable topological edge states in ABCA Graphene.} (\textbf{A}-\textbf{B}) STS LDOS map cuts at -20 meV (e) (within the overlapping gap of ABAB and ABCA graphene) and -10 meV (f) (at the flat band peak). (\textbf{C}) Band structure of ABAB graphene at a displacement field of 0.8 V/nm as calculated by DFT with self-consistent charge screening. The size of the blue circles is proportional to the projection of the wavefunction on the top layer. (\textbf{D}-\textbf{E}) STS LDOS map cuts (from same map as (A-B)) at 22.5 meV and 37.5 meV. (\textbf{F}) 2D Fourier transform of (D-E) (right) and a linecut through the primary electron scattering peaks (left). (\textbf{G}) Tight-binding LDOS profiles at fixed spatial intervals crossing an ABAB - ABCA domain wall with their corresponding distances (with respect to the domain boundary) color coded and written for the beginning and end point as well as the dominant in-gap topological edge state. (\textbf{H}) STS LDOS profiles at fixed spatial intervals crossing the ABAB - ABCA domain wall at the represented and color encoded distances (with respect to the domain wall as determind at higher energies) showing the dominant in gap topological edge state about 10 nm into the ABCA domain, consistent with tight binding in (G). STS measurements in (A-B), (D-E) and (G) were taken at setpoints of 400 mV and 150 pA with lock-in oscillations of 2.5 mV (A-B,D-E) and 1 mV (H).} 
	\end{figure*} 
	In order to understand the nature of the emergent insulating state, we theoretically investigate the ordered states that arise from adding correlation effects in this regime. To this end, we use a tight binding model with a Hamiltonian of $H=H_0+H_U$ with $H_0=\sum_{\vec k,a,b,\sigma} \epsilon_{\vec k,a,b}c^\dagger_{\vec k,a,\sigma}c_{\vec k,b,\sigma}$, the non-interacting part, and treat electron-electron interactions described by $H_U$ within a mean field approximation (see S3). We first concentrate on short-range Hubbard-type interactions between the up and down spin electrons of the material within each layer by including $H_U=U \sum_i\sum_a \left(n_{i,a,\uparrow}-\frac{1}{2}\right)\left(n_{i,a,\downarrow}-\frac{1}{2}\right)$  where $i$ runs over all lattice sites and $n_{i,a,\sigma}=c^\dagger_{i,a,\sigma}c_{i,a,\sigma}$ is the density at site $i$. We treat the term $H_U$ in a self-consistent mean-field decoupling yielding the results shown by the magenta curve in Figure 3D. The correlated (mean-field) state in this case is a Ferrimagnetic state within the topmost layer which spontaneously breaks the SU(2) invariance of the system and an antiferromagnetic ordering across the topmost and bottommost layer (opposite Ferrimagnetic state in the bottommost layer) as shown in Figure 3E. To have a more complete model, we now add longer range interactions across topmost to bottommost layers on A and B sites, respectively. This is a valid choice of the next order interactions as we know (from both tight-binding and DFT) that the top and bottom layer dominate the low-energy physics. Towards this end, we add an additional contribution, $U' \sum_i\sum_{\sigma,\sigma'} \left(n_{i,1A,\sigma}-\frac{1}{2}\right)\left(n_{i,4B,\sigma'}-\frac{1}{2}\right)$ to $H_U$ and study the ordered states in the self-consistent mean field approximation as we sweep $U$ and $U'$, the intralayer interactions and the top-bottom layer interactions. When turning on $U'$, at values $U' \approx U$, the ferrimagnetic state discussed above (at $U'$ = 0) gives way to a charge transfer (excitonic) insulating state which spontaneously breaks the inversion symmetry between bottom and top layer, spontaneously transferring charge across the A sites of the topmost and B sites of the bottommost layer as shown in Figure 3F. Due to the spontaneous charge transfer, the local density of states at the A sites of the topmost layer becomes electron-hole asymmetric in contrast to the local density of states in a ferrimagnetically ordered state (also shown in Figure 3E) that maintains electron-hole symmetry. We calculate a full phase diagram of the system as a function of $U$ and $U'$, shown in Figure 3G. The color scale on this plot displays the electron-hole asymmetry, which is seen to be generically large in the charge transfer excitonic insulator state. Based upon this criterion alone, our experimental spectra at most gate voltages match the expectation for the charge transfer excitonic insulator state. The only exception is the region around where the gap is a minimum, where it is observed to be more electron-hole symmetric.  While the electron-hole asymmetry gives us some insight into the nature of the correlated gap, future measurements at high magnetic field can definitively answer whether the state involves magnetic order.

	We next look at the consequence of the topological character of ABCA graphene in an applied displacement field. In Figure 4A, we show an STS LDOS image of several ABCA and ABAB domains at an electric field of 0.475 V/nm and -20 meV in energy. We clearly see the presence of edge states near ABCA-ABAB domain walls dominate the LDOS. Figure 4B (at the flat band peak, -10 meV) and D-E (above the flat band) show more energy slices from the same LDOS map as Figure 4A. In these images, there is no signature of the edge states, which only appear at particular energies and gate voltages. Towards understanding these edge states, we consider the nature of the ABAB domains in comparison to the ABCA domains. In Figure 4C we show the DFT calculated fat band structure of ABAB graphene under a displacement field (with the projection on the top layer shown by the size of the blue circles). The band structure is highly layer polarized. While the low energy band labeled 1 dominates the top layer, band 2 has nearly zero weight. Under the application of a displacement field, a gap is observed on the top layer (between bands 1 and 3), although the four layer structure as a whole remains metallic. This is also confirmed by STS measurements in the ABAB regions under a displacement field, Figure S4. We further confirm this band structure with quasiparticle interference (QPI) mapping. Figures 4D-E show STS LDOS map cuts at two energies in the conduction band of ABAB graphene. The wave-like structure in the ABAB domains originates from back-scattering from the ABCA domain walls. Shown in Figure 4F are 2D Fourier transforms (right) of the LDOS maps in Figs. 4D-E revealing the dominant hexagonal peaks from the scattering pattern. A line cut through the peaks (Figure 4F left) gives a direct visualization of the top layer quasi-parabolic band as well as a faint signature of the valence band (it dominates on the bottom layer), matching the theoretical band structure under field. Having confirmed the theoretically predicted band structure of ABAB graphene under a displacement field, we now consider its topological properties. An analysis of band 1 yields a valley Chern number of 1, with opposing signs in the two valleys (Figure S2). Thus, when considering the interface between ABCA and ABAB graphene seen in STM, both develop gaps in the presence of a displacement field with a difference in Chern number of 1 for each valley. This in turn implies that there are helical edge states of topological origin at the interface with one mode per valley (in contrast to the interface between AB and BA bilayer graphene where two modes per valley are expected \cite{huang18,ju15,vaezi13}). The dominating edges in the LDOS map of Figure 4A are a direct manifestation of these topological surface helical edge states.
	
	For further insight into the observed interface properties of this topological edge state, we study a finite size spatial region that includes a sharp boundary between ABCA and ABAB graphene at similar conditions as in experiment using tight-binding (more details in supplementary materials). In Figure 4G we show the results of these calculations for the LDOS at a number of positions across the ABCA-ABAB interface. The calculations show that the topological edge mode is localized around 8 nm away from the interface inside the ABCA domain, characterized by an in-gap state within the overlapping displacement field gaps of ABCA and ABAB graphene. To compare to theory, we show the corresponding experimentally measured LDOS spectra across the interface in Figure 4H. We see remarkable agreement between theory and experiment for the spatial LDOS dependence. Within the overlapping displacement field gaps of the LDOS, an in-gap state is present which in experiment is located 10 nm away from the domain boundary inside the ABCA region, in close accord with theory. These measurements establish that the interface of ABCA - ABAB domains (which can be easily realized by small angle tDBG) feature surface helical topological edge states which can be turned on and off with displacement field and gate voltage.
	
	Our experiments show that small angle tDBG moir\'e patterns are an easy platform to create large stable four-layer rhombohedral graphene domains. We commonly observe rhombohedral patches that are hundreds of nanometers to a micron in size, making them suitable for any technique including transport where electrodes can be directly placed on the rhombohedral regions. Stacking more bilayers at small angles may lead to even thicker rhombohedral graphene domains. Spectroscopy on ABCA graphene reveals a flat band of 3-5 meV half-width and a correlated gap at neutrality making it a unique correlated van der Waals system where there is no need for a moir\'e superlattice to modify the low energy electronic properties. Further exploration of ABCA graphene can provide answers to certain fundamental questions, for instance regarding the importance of moir\'e potentials and integer fillings in generating emergent phases such as superconductivity in neighboring regions of the phase diagram. The chiral nature of the tunable topological edge states in rhombohedral graphene also makes it of great interest to observe gyrotropic effects and towards Floquet engineering. 
	
	
	\section*{Acknowledgments}
	This work was supported by Programmable Quantum Materials, an Energy Frontier Research Center funded by the U.S. Department of Energy (DOE), Office of Science, Basic Energy Sciences (BES), under award DE-SC0019443. STM equipment support was provided by the Air Force Office of Scientific Research via grant FA9550-16-1-0601 and by the Office of Naval Research via grant N00014-17-1-2967. DMK acknowledges funding by the Deutsche Forschungsgemeinschaft (DFG, German Research Foundation) under Germany's Excellence Strategy - Cluster of Excellence Matter and Light for Quantum Computing (ML4Q) EXC 2004/1 - 390534769. Gef\"ordert durch die Deutsche Forschungsgemeinschaft (DFG) im Rahmen der Exzellenzstrategie des Bundes und der L\"ander - Exzellenzcluster Materie und Licht f\"ur Quanteninformation (ML4Q) EXC 2004/1 - 390534769. LX and AR acknowledge the support by the European Research Council (ERC-2015-AdG694097), cluster of Excellence AIM, SFB925 and Grupos Consolidados (IT1249-19). Growth of hexagonal boron nitride crystals was supported by the Elemental Strategy Initiative conducted by the MEXT, Japan and the CREST (JPMJCR15F3), JST. The Flatiron Institute is a division of the Simons Foundation. We acknowledge support from the Max Planck—New York City Center for Non-Equilibrium Quantum Phenomena.

	\bibliographystyle{unsrtnat}
	\bibliography{tdbgbibtex}
		
\end{document}


\setcounter{equation}{0}
		\setcounter{figure}{0}
		\setcounter{table}{0}
		\setcounter{page}{1}
		\setcounter{section}{0}
		\renewcommand{\theequation}{S\arabic{equation}}
		\renewcommand{\thefigure}{S\arabic{figure}}
	\title{Supplementary Materials for \\ Moir\'e-less Correlations in ABCA Graphene}

\affiliation{
	Department of Physics, Columbia University, New York, New York 10027, United States\looseness=-1}
\affiliation{
	Max Planck Institute for the Structure and Dynamics of Matter, Luruper Chaussee 149, 22761 Hamburg, Germany\looseness=-1}
\affiliation{
	Institut fur Theorie der Statistischen Physik, RWTH Aachen University, 52056 Aachen, Germany and JARA-Fundamentals of Future Information Technology, 52056 Aachen, Germany\looseness=-2}
\affiliation{
	Department of Mechanical Engineering, Columbia University, New York, NY, USA\looseness=-1}
\affiliation{
	National Institute for Materials Science, 1-1 Namiki, Tsukuba 305-0044, Japan\looseness=-1}
\affiliation{
	Center for Computational Quantum Physics (CCQ), The Flatiron Institute, 162 Fifth Avenue, New York, NY 10010, USA\looseness=-1}
\affiliation{
	Nano-Bio Spectroscopy Group, Departamento de Fisica de Materiales, Universidad del País Vasco, 20018 San Sebastian, Spain\looseness=-1}

\author{Alexander Kerelsky}
\altaffiliation{These authors contributed equally to this work.
}
\affiliation{
	Department of Physics, Columbia University, New York, New York 10027, United States\looseness=-1}
\author{Carmen Rubio-Verd\'u}
\altaffiliation{These authors contributed equally to this work.
}
\affiliation{
	Department of Physics, Columbia University, New York, New York 10027, United States\looseness=-1}
\author{Lede Xian}
\affiliation{
	Max Planck Institute for the Structure and Dynamics of Matter, Luruper Chaussee 149, 22761 Hamburg, Germany\looseness=-1}
\author{Dante M. Kennes}
\affiliation{
	Institut fur Theorie der Statistischen Physik, RWTH Aachen University, 52056 Aachen, Germany and JARA-Fundamentals of Future Information Technology, 52056 Aachen, Germany\looseness=-2}
\author{Dorri Halbertal}
\affiliation{
	Department of Physics, Columbia University, New York, New York 10027, United States\looseness=-1}
\author{Nathan Finney}
\affiliation{
	Department of Mechanical Engineering, Columbia University, New York, NY, 		USA\looseness=-1}
\author{Larry Song}
\affiliation{
	Department of Physics, Columbia University, New York, New York 10027, United States\looseness=-1}
\author{Simon Turkel}
\affiliation{
	Department of Physics, Columbia University, New York, New York 10027, United States\looseness=-1}
\author{Lei Wang}
\affiliation{
	Department of Physics, Columbia University, New York, New York 10027, United States\looseness=-1}
\author{K. Watanabe}
\affiliation{
	National Institute for Materials Science, 1-1 Namiki, Tsukuba 305-0044, Japan\looseness=-1}
\author{T. Taniguchi}
\affiliation{
	National Institute for Materials Science, 1-1 Namiki, Tsukuba 305-0044, Japan\looseness=-1}
\author{James Hone}
\affiliation{
	Department of Mechanical Engineering, Columbia University, New York, NY, USA\looseness=-1}
\author{Cory Dean}
\affiliation{
	Department of Physics, Columbia University, New York, New York 10027, United States\looseness=-1}
\author{Dmitri Basov}
\affiliation{
	Department of Physics, Columbia University, New York, New York 10027, United States\looseness=-1}
\author{Angel Rubio}
\altaffiliation{Correspondence to:
	\href{mailto:apn2108@columbia.edu}{apn2108@columbia.edu} (A.N.P); \\
	\href{mailto:angel.rubio@mpsd.mpg.de}{angel.rubio@mpsd.mpg.de} (A.R.)}
\affiliation{
	Max Planck Institute for the Structure and Dynamics of Matter, Luruper Chaussee 149, 22761 Hamburg, Germany\looseness=-1}
\affiliation{
	Center for Computational Quantum Physics (CCQ), The Flatiron Institute, 162 Fifth Avenue, New York, NY 10010, USA\looseness=-1}
\affiliation{
	Nano-Bio Spectroscopy Group, Departamento de Fisica de Materiales, Universidad del País Vasco, 20018 San Sebastian, Spain\looseness=-1}
\author{Abhay N. Pasupathy}
\altaffiliation{Correspondence to:
	\href{mailto:apn2108@columbia.edu}{apn2108@columbia.edu} (A.N.P); \\
	\href{mailto:angel.rubio@mpsd.mpg.de}{angel.rubio@mpsd.mpg.de} (A.R.)}
\affiliation{
	Department of Physics, Columbia University, New York, New York 10027, United States\looseness=-1}

	\date{\today}
	
	\maketitle

	\onecolumngrid
	
	\setlength{\parindent}{0cm}
	\textbf{This PDF file includes:}\\
	Methods\\
	Tiny Angle Twisted Bilayer Graphene\\
	Berry Curvature in ABCA and ABAB Graphene\\
	Mean-Field Treatment of the Correlation\\
	ABAB Field Dependent STS\\
	Figures S1-S10\\
	
	\setlength{\parindent}{24pt}

	\section*{Methods: Experiment}
	Our fabrication of twisted double bilayer graphene (tDBG) samples follows the established “tearing” method, using PPC as a polymer to sequentially pick up hBN, half of a piece of graphene followed by the second half with a twist angle. This structure is flipped over and placed on an Si/SiO$_2$ chip. Directly contact is made to the tDBG via $\mu$soldering with Field’s metal\cite{girit08}, keeping temperatures below 80 C during the entire process to minimize the chance of layers fully rotating back to Bernal stacking which happens on annealing the structures.
	
	STM measurements are taken in a home-built 5.7 Kelvin system at ultra high vacuum. STM tips are prepared and calibrated for atomic sharpness and electronic integrity on freshly prepared Au (111) crystals. Samples were measured with multiple tips to ensure consistency of results.
	
	Nano-infrared imaging in this work was performed using a commercial (Neaspec) scattering-type scanning near-field optical microscope (s-SNOM). In this measurement, mid-IR light (continuous wave CO$_2$ gas laser at a wavelength of 10.6 $\mu$m) was focused on the apex of a metallic tip. The scattered light was collected by a cryogenic HgCdTe detector (Kolmar Technologies). The AFM tip was excited at a frequency of about 75 kHz, with a tapping amplitude of about 60 nm. The far-field contribution to the signal can be eliminated from the signal by locking to a high harmonic (here we used the 3rd harmonic of the tapping). The phase of the backscattered signal was extracted using an interferometric detection method, the pseudo-heterodyne scheme, by interfering at the detector the scattered light with a modulated reference arm.
	
	\section*{Methods: Tight Binding and Density Functional Theory}
	The band structures and layer-projected local density of states (LDOS) of multilayer graphene with ABCA and ABAB stacking are calculated with first-principle calculations based on density functional theory as implemented in the Vienna Ab initio Simulation Package (VASP)\cite{kresse}.  The pseudopotentials are constructed with the projector augmented wave method (PAW)\cite{blochl}.  The exchange-correlation functionals are treated at the local-density approximations (LDA)\cite{perdeww} level. Plane-wave basis sets are employed with an energy cutoff of 600 eV. A dense k-point grid of 150x150x1 is adopted in the self-consistent calculations to ensure the convergence of the band structures and the LDOS. A vacuum region larger than 15 Angstrom is added along the z-direction to eliminate artificial interactions between slab images.  Dipole corrections are included in all the calculations with displacement fields.  
	
	To simulate the STS measurement across the ABCA-ABAB domain walls, we calculate the LDOS at different locations across the domain walls with tight-binding model calculations. We model the domain walls with a flake of four-layer graphene with area larger than 200 nm x 200 nm in the real space.  In the region with x $<$ -1 nm, the stacking sequence is ABCA. In the region between x = -1nm and x = 1nm, the x coordinates of all lattice sites in the top two layers are linearly expanded such that the stackings become ABAB in the region with x $>$1 nm.  We model the system with the following tight-binding Hamiltonian\cite{tram10}: $\mathbf{H}=\sum_{i j} t_{i j}|i\rangle\langle j|$
	where t$_{ij}$ is the hopping parameter between pz orbitals at the two lattice sites r$_i$ and r$_j$ and it has the following form: $	t_{i j}=n^{2} \gamma_{0} \exp \left[\lambda_{1}\left(1-\frac{\left|\boldsymbol{r}_{i}-\boldsymbol{r}_{j}\right|}{a}\right)\right]+\left(1-n^{2}\right) \gamma_{1} \exp \left[\lambda_{2}\left(1-\frac{\left|\boldsymbol{r}_{i}-\boldsymbol{r}_{j}\right|}{c}\right)\right]$
	where a=1.412 $\AA$ is the in plane C-C bond length, c=3.36 $\AA$ is the interlayer separation, n is the direction cosine of r$_i$-r$_j$ along the out of plane axis (z-axis), $\gamma_0$ ($\gamma_1$) is the intralayer (interlayer) hopping parameter, and $\lambda_1$ ($\lambda_2$) is the intralayer (interlayer) decay constant. For the interlayer interaction, only the hopping between atoms in adjacent layers are considered.  This tight-binding model has been shown to reproduce the low-energy structure of graphene calculated by local density functional theory (DFT) calculations with the following value for the parameters: $\gamma_0$ = -2.7 eV, $\gamma_1$ = 0.48 eV, $\lambda_1$ = 3.15 and $\lambda_2$ = 7.50 \cite{li10}. For the calculation of local density of states, we employ the Lanczos recursive method\cite{wang12}.

	\section*{Tiny Angle Twisted Bilayer Graphene}
	
	For comparison to small angle twisted double bilayer graphene (tDBG), in Figure S1 we show a small angle twisted bilayer graphene (tBG) STM topography. It is clear from the topography and main text Figure 1D that the contrast is fundamentally different in small angle tDBG as opposed to tBG. The main contrast in tBG emerges from the AA sites and the SP domain walls. In between, there are maximized (in space) commensurate Bernal stacked AB and BA regions which show the same contrast. On the other hand in tDBG the dominant contrast comes from two sets of different commensurate stacking domains -- ABAB (Bernal) and ABCA (rhombohedral) graphene as described in the text.
	
	\begin{figure*}[h]
		\includegraphics[width=\linewidth]
		{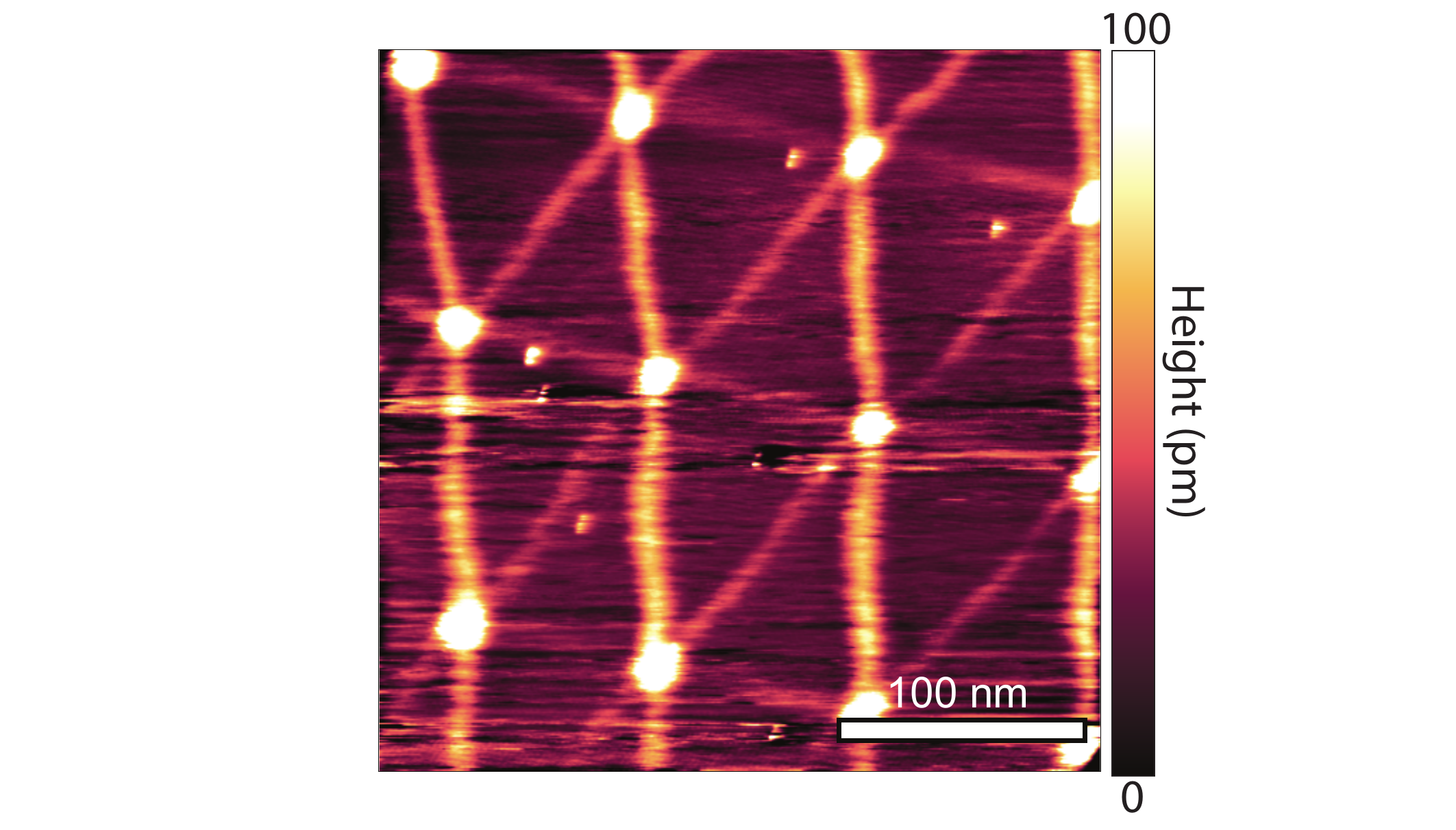}
		\caption{
			A small angle twisted bilayer graphene (tBG) moir\'e. The dominant contrast for small angle tBG comes from the AA sites and the SP domain walls. The AB versus BA commensurate stacking domains show the same contrast unlike the ABAB versus ABCA commensurate stacking domains in small angle tDBG.} 
	\end{figure*} 
			
	\section*{Berry Curvature in ABCA and ABAB Graphene}
	In order to investigate the topological properties of ABCA and ABAB graphene, we calculate the berry curvatures for the low energy bands for the two systems under applied displacement field. We use the tight-binding model calculations with hopping between lattice sites as given in the methods section. To simulate the applied displacement field, the onsite potential difference between adjacent layers is set to be 0.08 eV. Figure S2A shows the Berry curvature calculated for the bottom of the conduction band of ABCA graphene for half of the Brillouin zone centered around the K point. It can be seen that there is a large peak around the K point. If we integrate the Berry curvature around the region close to this K valley, we can obtain a valley Chern number of +2 for the bottom of the conduction band of ABCA graphene. 
	
		\begin{figure*}[t]
		\includegraphics[width=\linewidth]
		{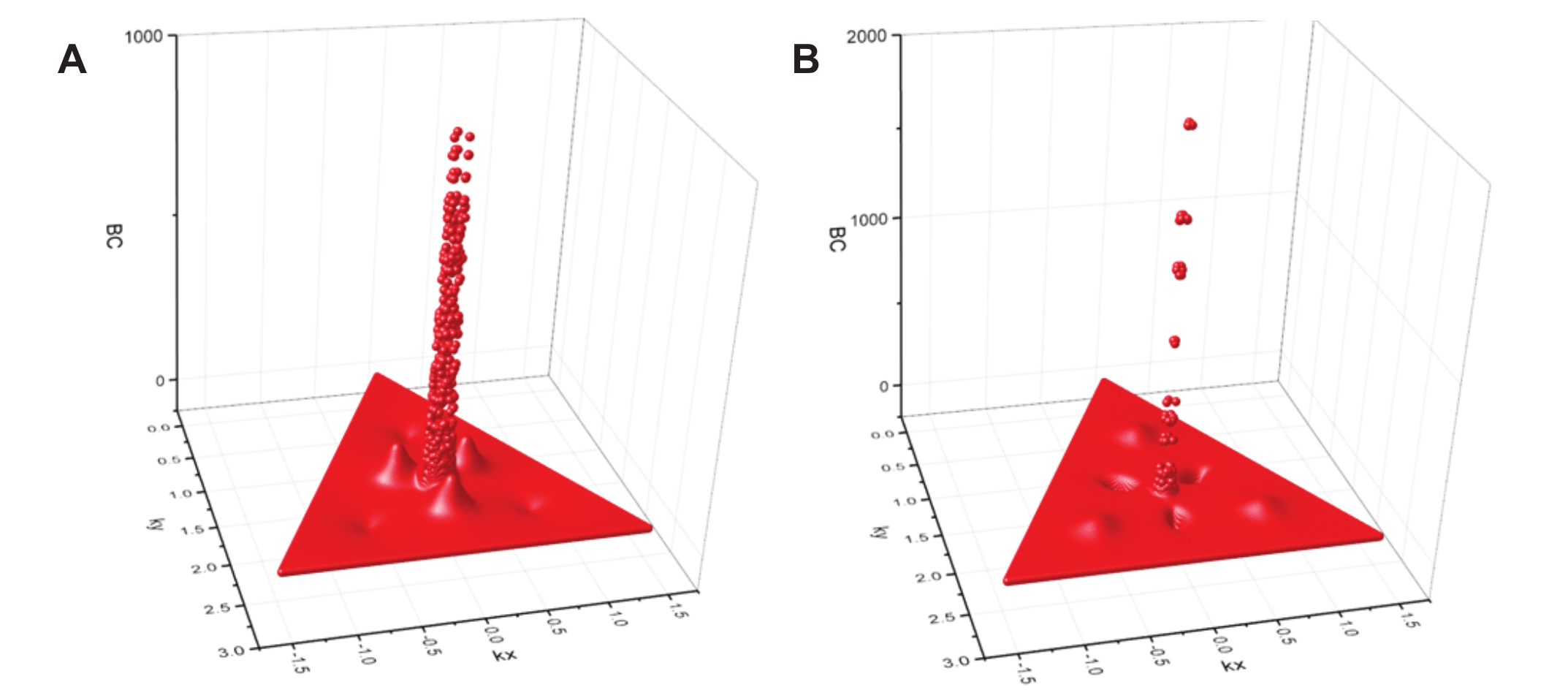}
		\caption{
			Berry curvature for the bottom of the ABCA conduction band (A) and ABAB conduction band (B)} 
	\end{figure*} 
	
	There is no overall band gap for ABAB graphene even under applied displacement field. However, if we just consider the states at the surface layer (the top or the bottom), there is indeed a band gap between surface bands 1 and 3 (see main text Figure 4C). The Berry curvature calculated for the bottom of conduction band 1 is shown in Figure S2B. If we integrate the Berry curvature around the region close to the K valley, we obtain a valley Chern number of +1, which is different from the case of ABCA graphene. The difference of the valley Chern number between the surface bands of ABCA and ABAB graphene implies there will be surface helical edge states at the interfaces of the two systems.

	\section*{Mean-field treatment of the correlation}
	
	Here we analyze the effect of correlations on the density of states at the A sites of the topmost layer. We use a tight-binding model with Hamiltonian $H=H_0+H_U$ with $H_0=\sum_{\vec k,a,b,\sigma} \epsilon_{\vec k,a,b}c^\dagger_{\vec k,a,\sigma}c_{\vec k,b,\sigma}$ the non-interacting part and treat the electron-electron interaction described by $H_U$ within a mean-field approximation. The indices $a$ and $b$ take values $1A,1B,2A,2B,3A,3B,4A,4B$, where the number indicates the layer (topmost layer: $1$, bottommost layer: $4$) and $A$ and $B$ label the two sites in the unit cell per layer. In the following we thus represent the dispersion as a $8\times8$ matrix labeling these different degrees of freedom. In agreement with the ab-initio characterization for the regions being ABCA stacked (main text) we consider
	\begin{equation}
	\epsilon_{\vec k}=\begin{pmatrix}
	-E_f/2& f(0,\vec k) & 0& 0& 0& 0&0&0\\
	f(0,\vec k)^*& -E_f/2 & t'& 0& 0& 0&0&0\\
	0&t'&	 -E_f/6 &f(1,\vec k)& 0& 0& 0&0\\
	0&0&	 f(1,\vec k)^* &-E_f/6& t'& 0& 0&0\\
	0&0&	 0 &t'&E_f/6& f(2,\vec k)& 0& 0\\
	0&0&	 0 &0& f(2,\vec k)&E_f/6& t'& 0\\
	0&0&	 0 &0&0&t'&E_f/2& f(3,\vec k)\\
	0&0&	 0 &0&0&0&f(3,\vec k)^*&E_f/2\\
	\end{pmatrix}
	\end{equation}
	
	while for the ABAB regions we choose
	
	\begin{equation}
	\epsilon_{\vec k}=\begin{pmatrix}
	-E_f/2& f(0,\vec k) & 0& 0& 0& 0&0&0\\
	f(0,\vec k)^*& -E_f/2 & t'& 0& 0& 0&0&0\\
	0&t'&	 -E_f/6 &f(1,\vec k)& 0& t'& 0&0\\
	0&0&	 f(1,\vec k)^* &-E_f/6& 0& 0& 0&0\\
	0&0&	 0 &0&E_f/6& f(2,\vec k)& 0& 0\\
	0&0&	 t' &0& f(2,\vec k)&E_f/6& t'& 0\\
	0&0&	 0 &0&0&t'&E_f/2& f(3,\vec k)\\
	0&0&	 0 &0&0&0&f(3,\vec k)^*&E_f/2\\
	\end{pmatrix}
	\end{equation}
	
	\begin{figure}[t]
		\centering
		\includegraphics[width=0.45\columnwidth]{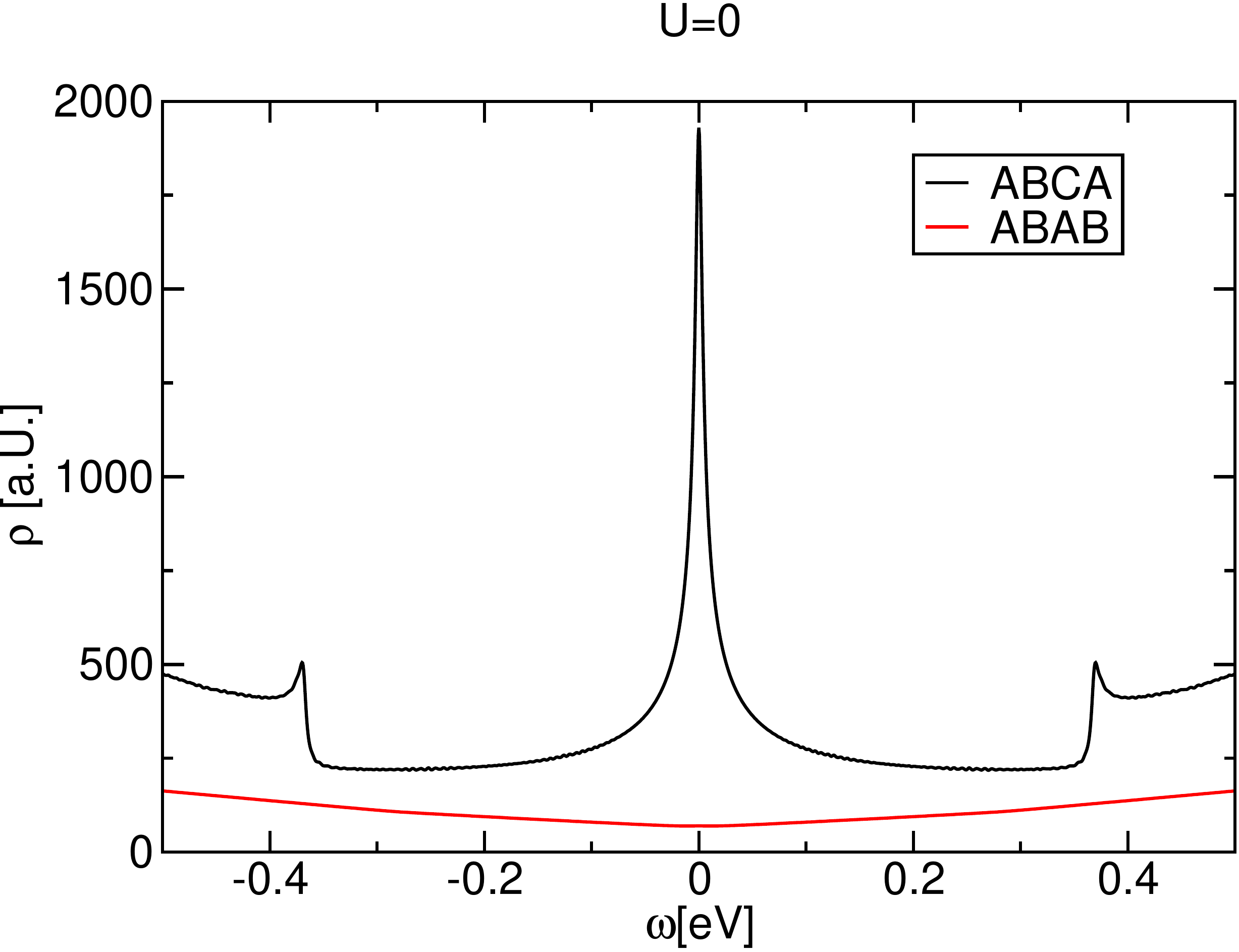}
		\hfill
		\includegraphics[width=0.45\columnwidth]{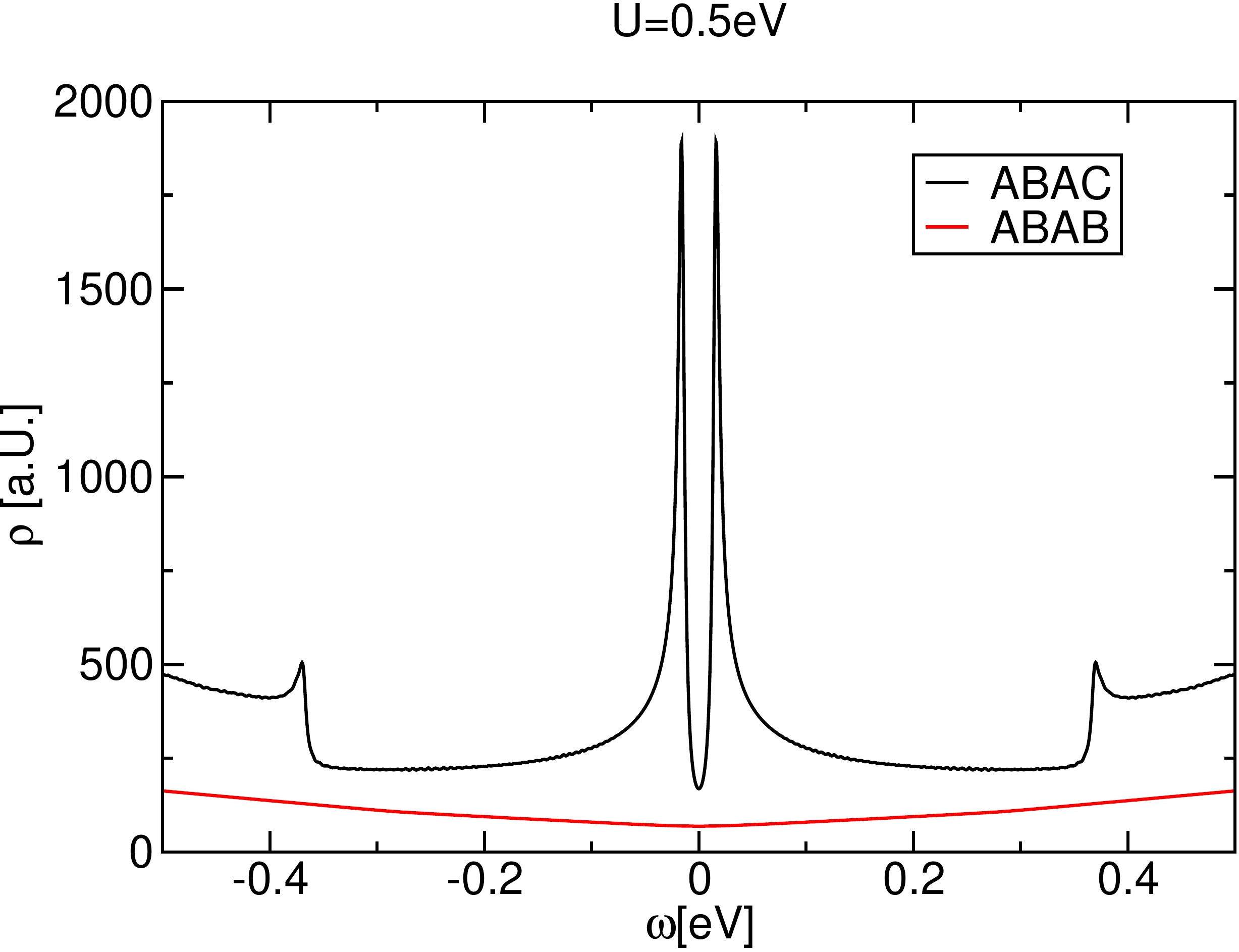}
		\caption{Left: Comparing the ABAB versus the ABCA non-interacting density of states at the A sites of the topmost layer. For the ABCA we find a square root divergence at low frequencies. Right: The same including local interactions $U$ within a mean field approach.}
		\label{fig:DOS_0}
	\end{figure}

	with $f(x,\vec k)=t \exp(i\;x\;\vec k\cdot\vec a)[\exp(i\vec k\cdot\vec a)+\exp(i\vec k\cdot\vec b)+\exp(i\vec k\cdot\vec c)]$, $t=-3.375eV$, $t'=0.46eV$, $\vec a=(1,0)$, $\vec b=(-1/2,\sqrt{3}/2)$, $\vec c=(-1/2,-\sqrt{3}/2)$ and where $E_f$ models an electric field applied across the layers.
	We neglect longer-ranged hoppings across layers for simplicity as they constitute a correction which does not influence the physics we discuss in the following.
	
	The non-interacting density of states at the A sites of the topmost layer is
	\begin{equation}
	\rho(\omega)\sim \int d^2k\;\; {\rm Im}\left[\frac{1}{\omega-\epsilon_{\vec k}+i\eta}\right]_{1A,1A}
	\end{equation}
	is shown for $E_f=0$ in the left panel of Fig.~\ref{fig:DOS_0} for $\eta=0.003$.

	\begin{figure}[t]
		\centering
		\includegraphics[width=\columnwidth]{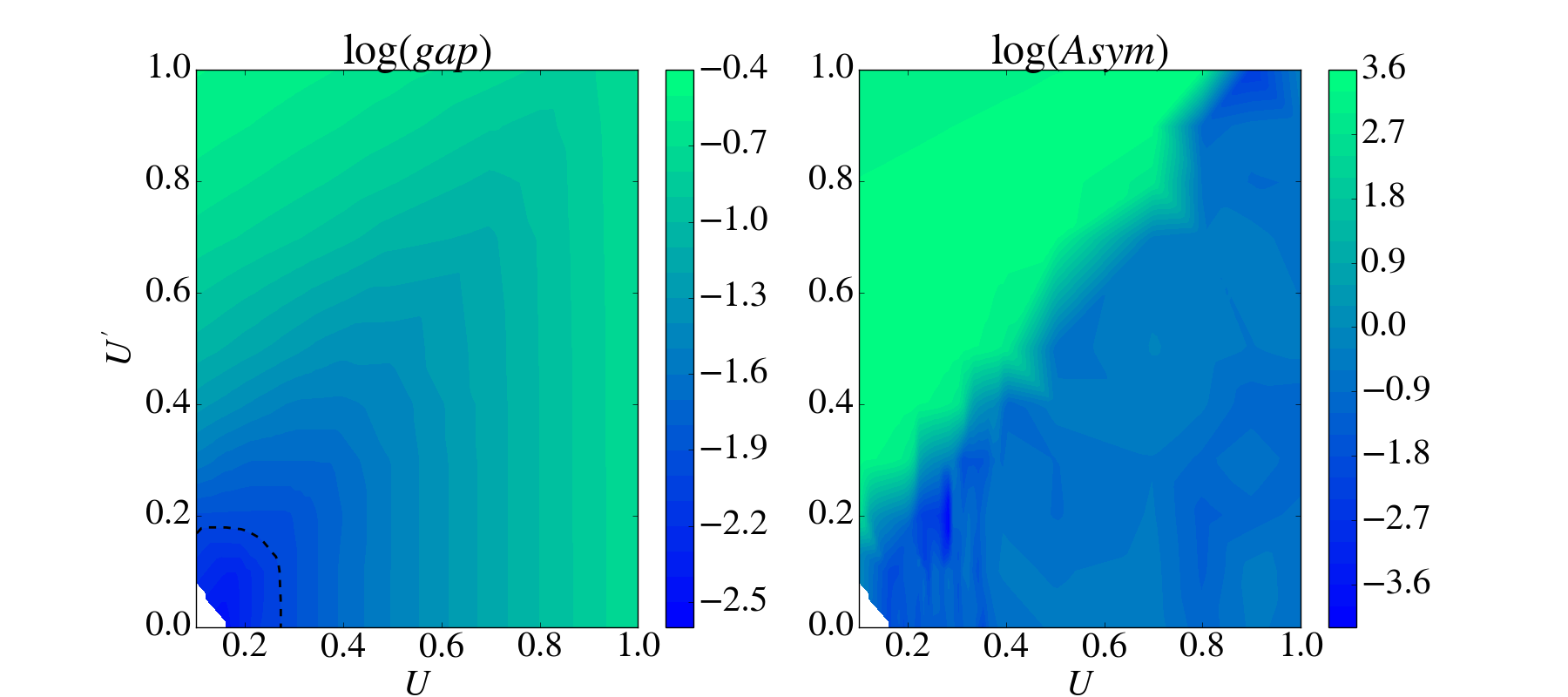}
		\caption{Characterization of order in the ABCA regions sweeping $U$ and $U'$. Left: logarithm of the gap size (in meV). The dashed line indicates a gap size of about $10$meV. Right: logarithm of the asymmetry of the two peaks in the correlation-split van-Hove singularity. The asymmetry indicates the charge transfer (excitonic) state.    }
		\label{fig:Sweep}
	\end{figure}

	\begin{figure}[t]
		\centering
		\includegraphics[width=.32\columnwidth]{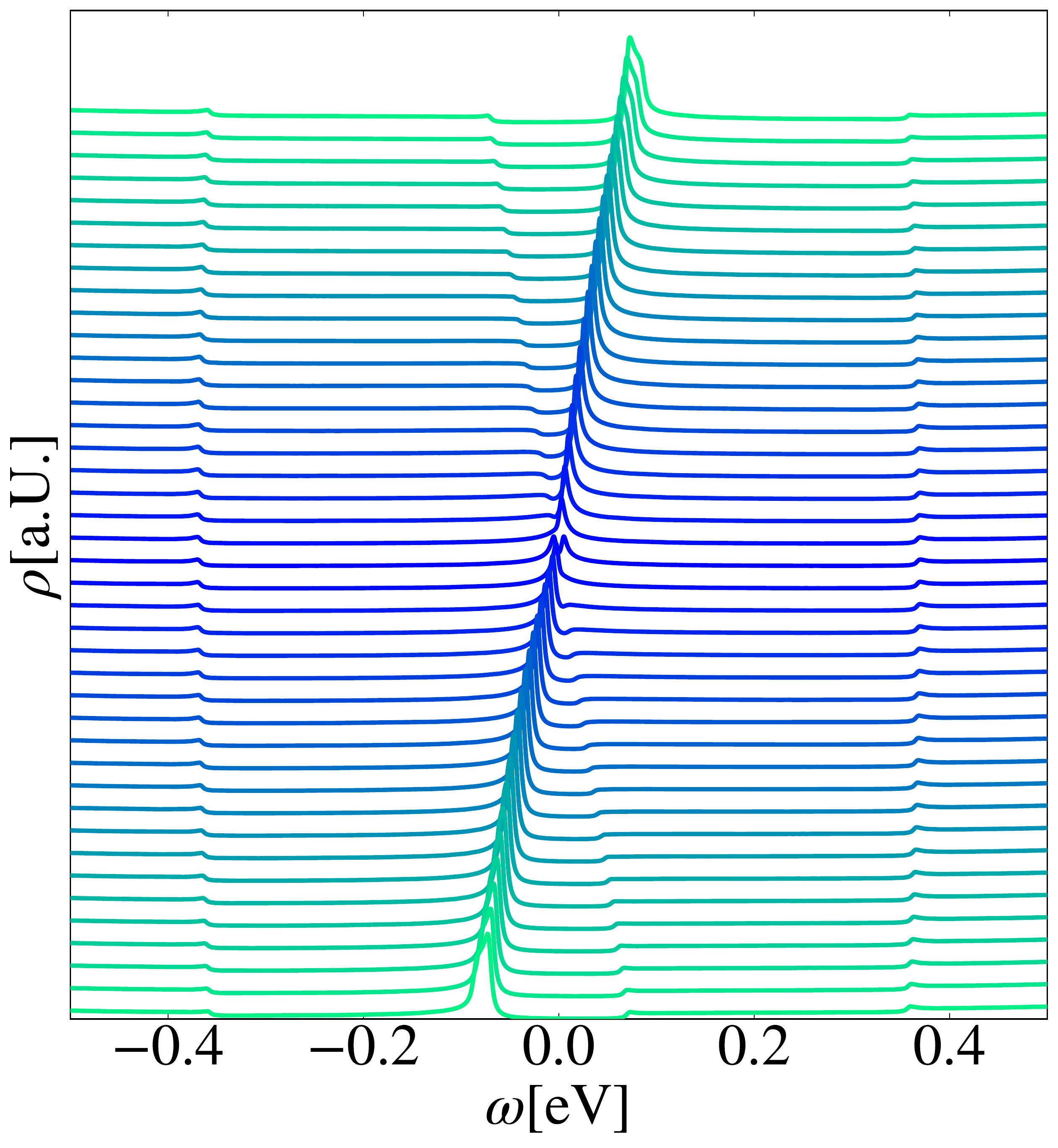}
		\hfill
		\includegraphics[width=.32\columnwidth]{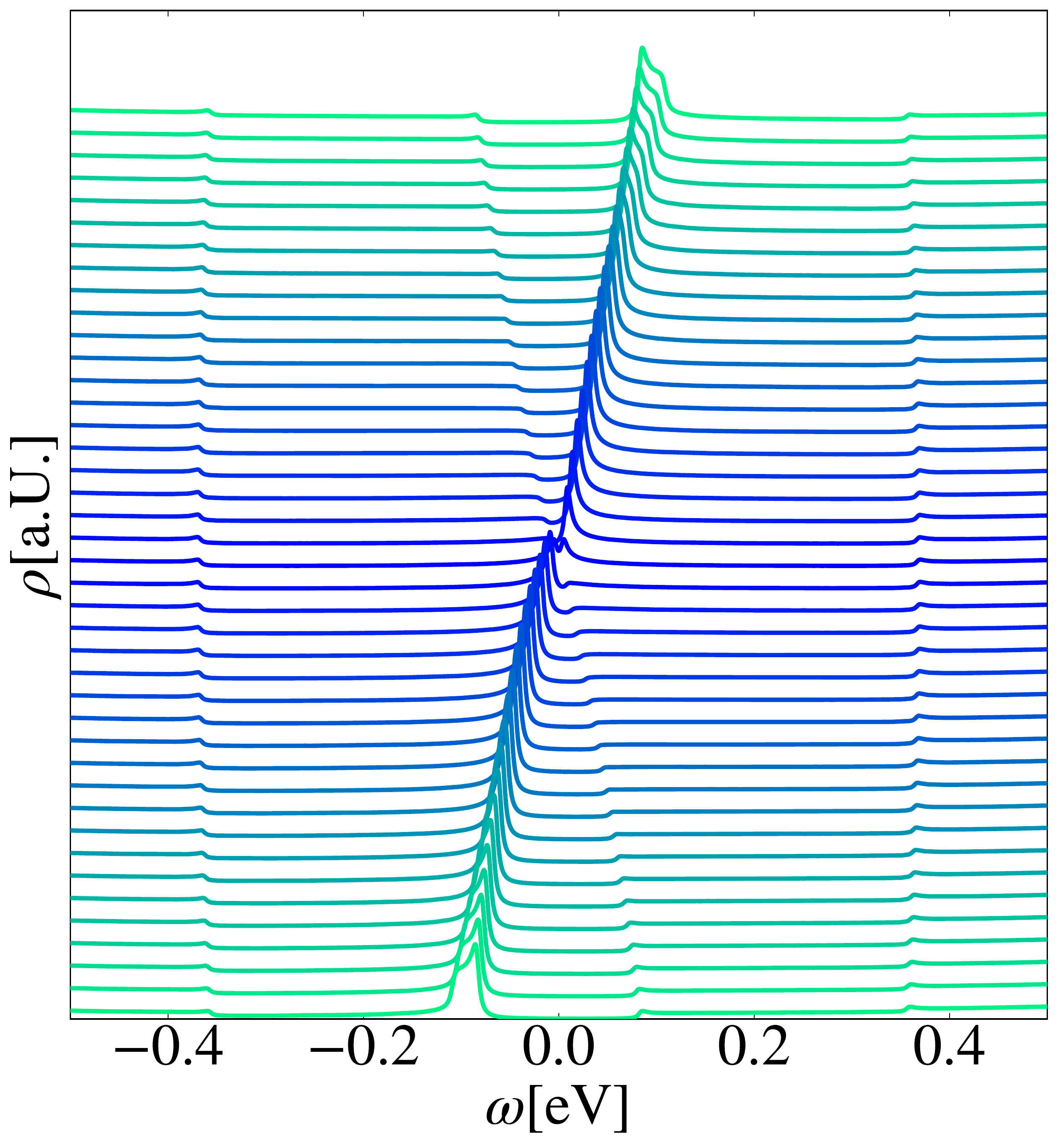}
		\hfill
		\includegraphics[width=.32\columnwidth]{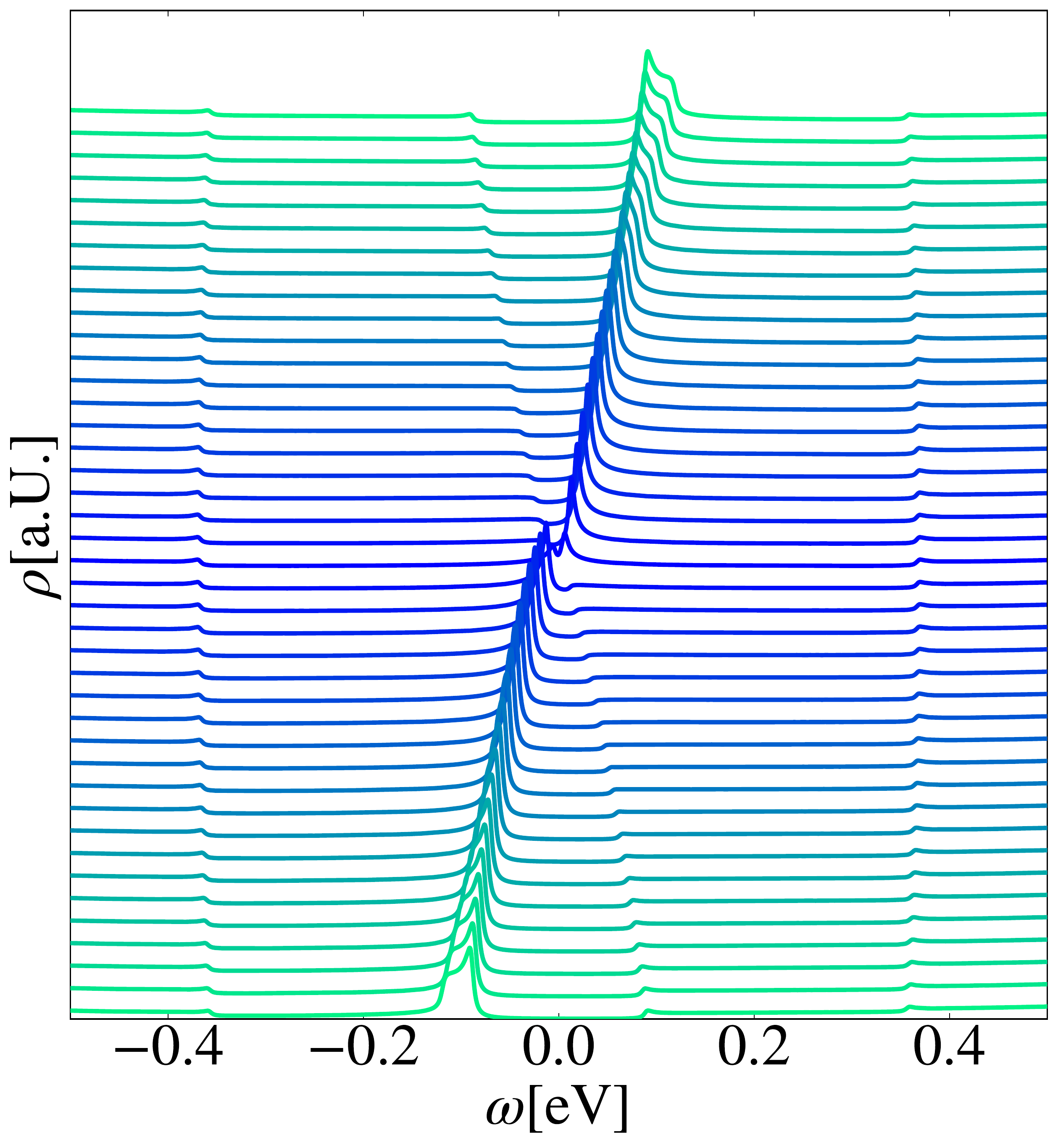}
		\caption{Waterfall plot of the local density of states including an Electric field sweept from $E_f=-0.2eV$ to $E_f=0.2eV$ in steps of $0.01eV$ for values which reproduce a van-Hove singularity splitting of about $10$meV at $E=0$. Left: $U=0.275eV$ and  $U'=0$. Center: $U=0.2V$ and  $U'=0.175$.  Right: $U=0.1V$ and  $U'=0.2$. $T=4K$ in all panels.   }
		\label{fig:DOS_E}
	\end{figure}

	\begin{figure}[t]
		\centering
		\includegraphics[width=.32\columnwidth]{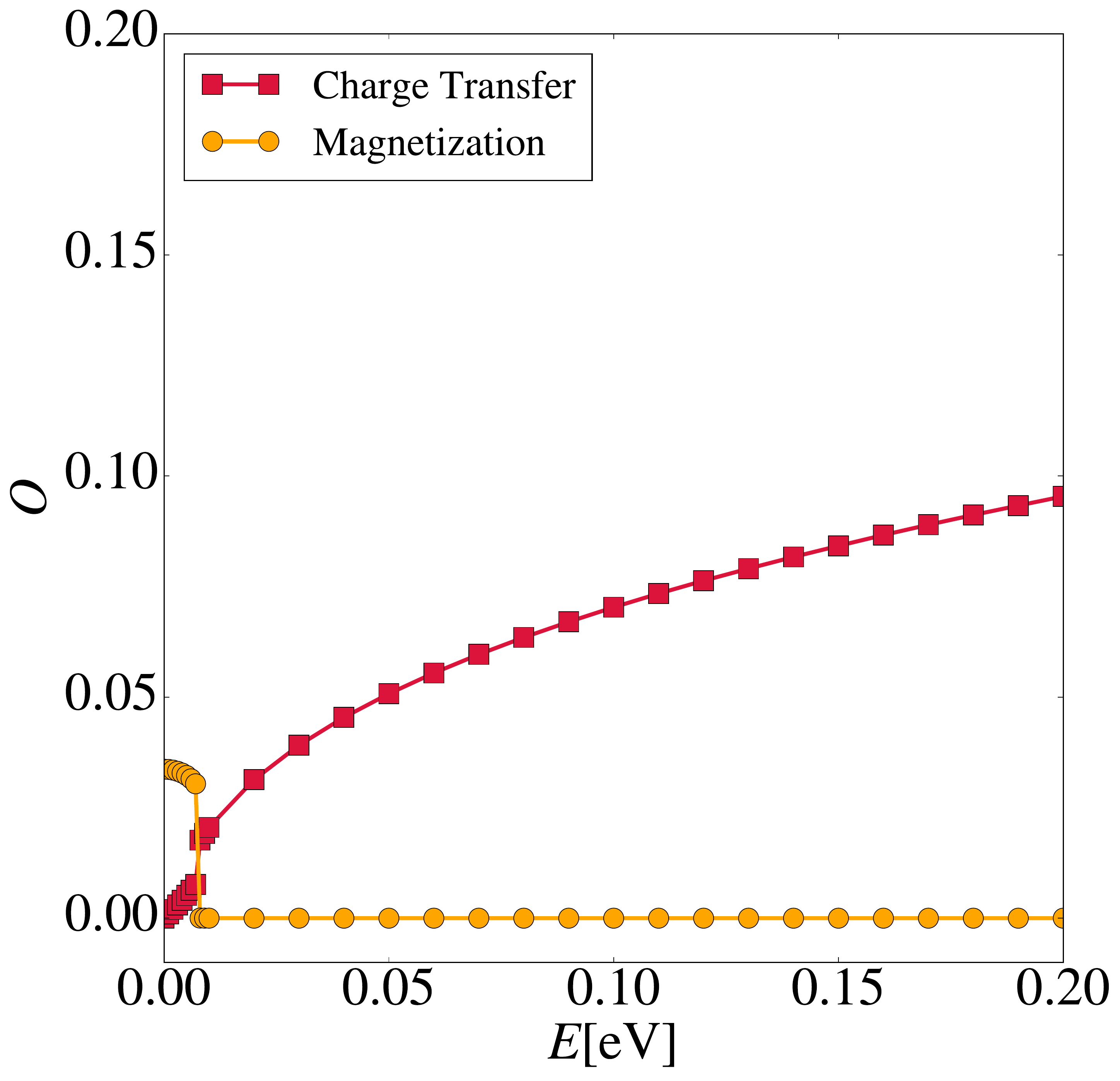}
		\hfill
		\includegraphics[width=.32\columnwidth]{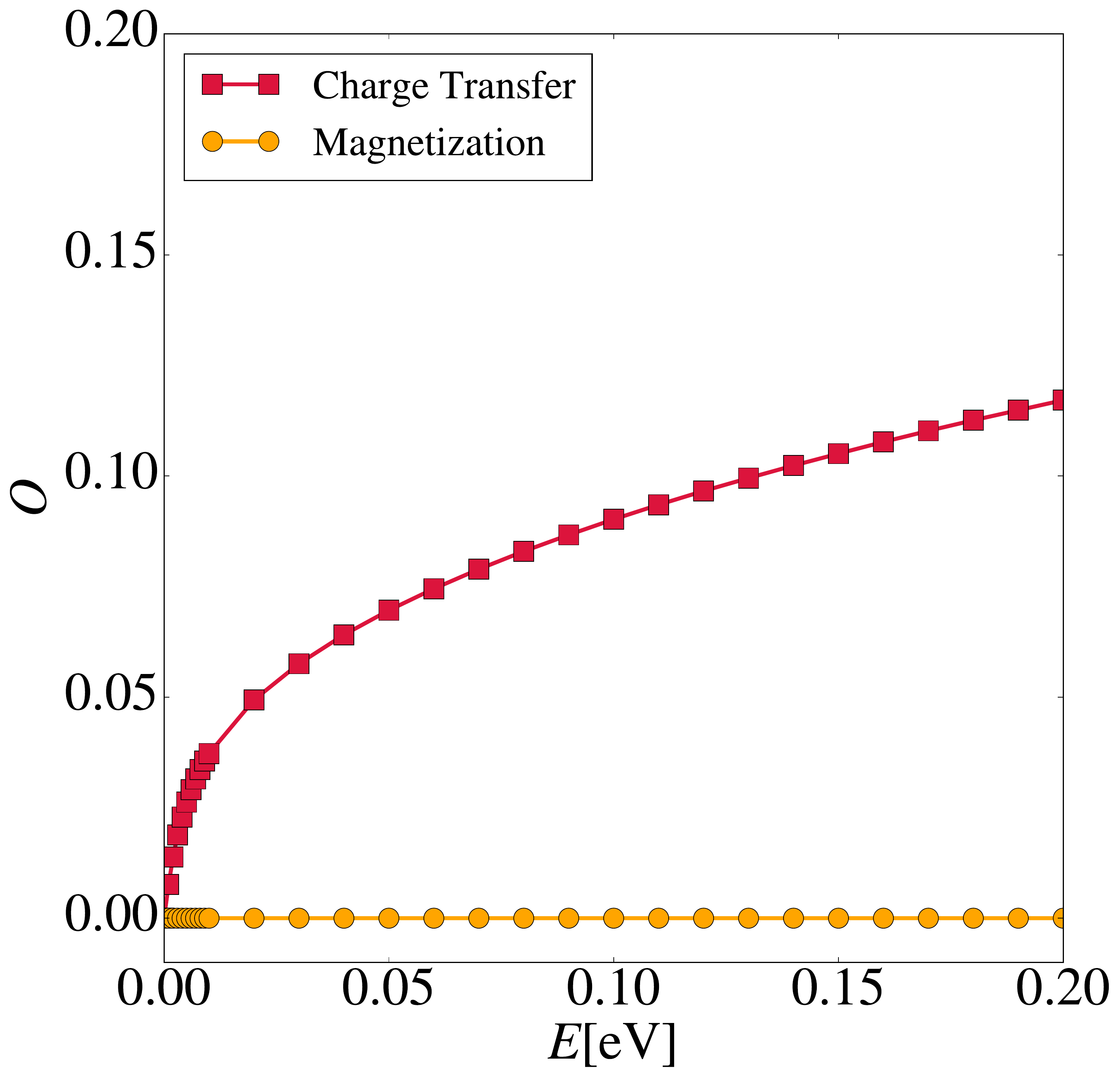}
		\hfill
		\includegraphics[width=.32\columnwidth]{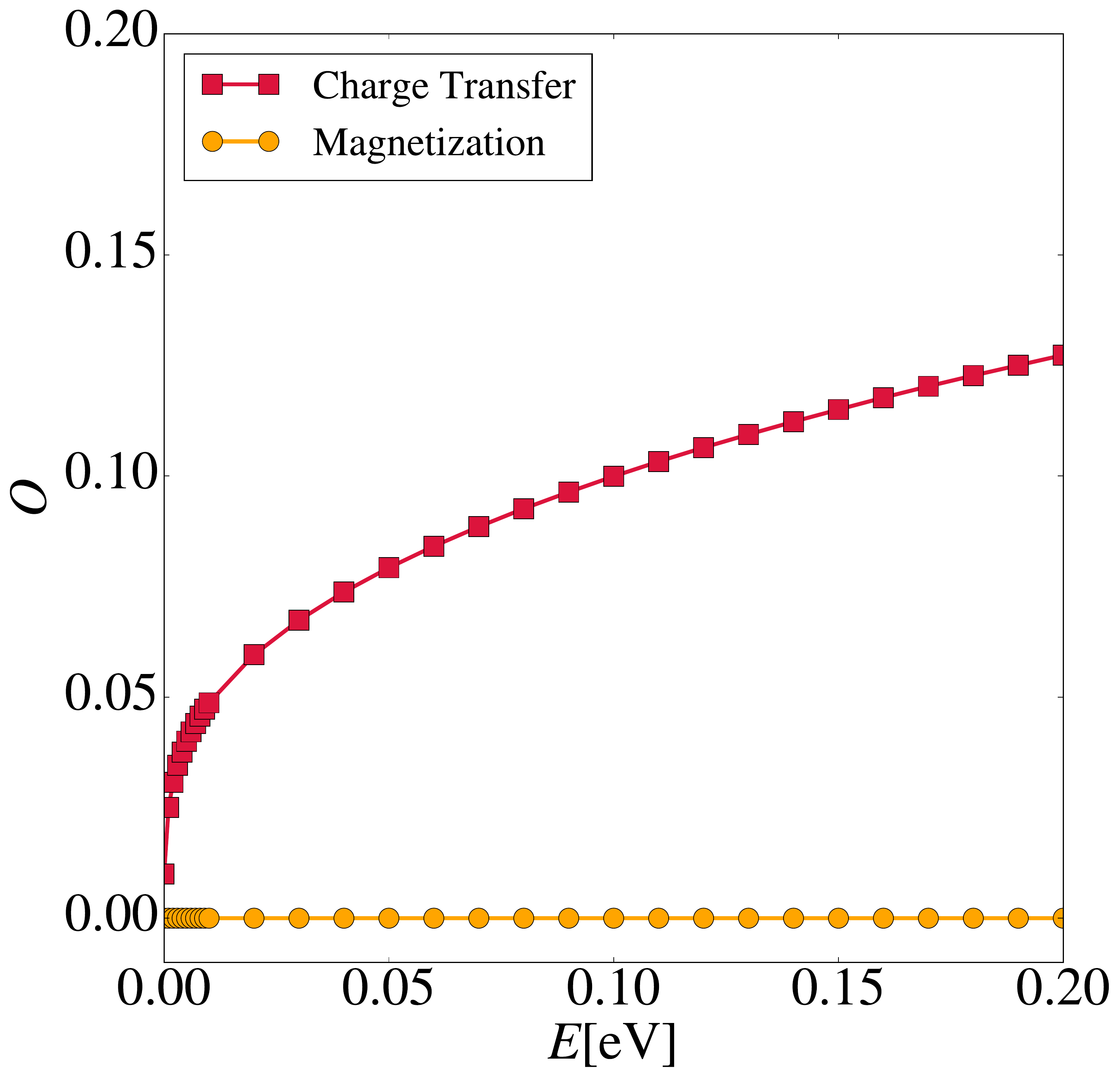}
		\caption{Order parameter for Ferrimagentic order as well as charge transfer order in dependence of applied electric field for values which reproduce a van-Hove singularity splitting of about $10$meV at $E_f=0$ and $T=4K$. Left: $U=0.275eV$ and  $U'=0$. Center: $U=0.2V$ and  $U'=0.175$.  Right: $U=0.1V$ and  $U'=0.2$.(right)   }
		\label{fig:SUS_E}
	\end{figure}

	In the case of ABAB stacking the four low energy bands touch at the Fermi-level in a quadratic fashion (two with positive and two with negative curvature). This gives rise to a constant density of states for $\omega\to 0$ in two dimensions. In contrast the ABCA stacking exhibits a quartic touching point of the two low energy  bands (one with positive and one with negative curvature) yielding a divergent density of states $\rho\sim 1/\sqrt{\omega}$ as  $\omega\to 0$. This diverging density of states will promote strong correlation effects in the sample. To illustrate this we fist concentrate on a short-ranged Hubbard-type interaction between the up and down spin electrons of the material by including 
	\begin{equation}
	H_U=U \sum_i\sum_a \left(n_{i,a,\uparrow}-\frac{1}{2}\right)\left(n_{i,a,\downarrow}-\frac{1}{2}\right)
	\end{equation}
	where $i$ runs over all lattice sites and $n_{i,a,\sigma}=c^\dagger_{i,a,\sigma}c_{i,a,\sigma}$ is the density at site $i$. The additional terms $-1/2$ are added for convenience and ensure that $\mu=0$ corresponds to the undoped case. We treat the term $H_U$ in a self-consistent mean-field decoupling and obtain the results shown in the right panel of Fig.~\ref{fig:DOS_0}  for $U=0.5eV$ and $T=4K$ (using again $\eta=0.003$). The effect on the local density of states is minute for the ABAB region, while a prominent splitting of the peaks appears for the ABCA stacking configuration. The correlated (mean-field) state in this case is a Ferrimagnetic state within the topmost layer which spontaneously breaks the $SU(2)$ invariance of the systems and an antiferromagnetic ordering across the topmost and bottommost layer (opposite Ferrimagentic state in the bottommost layer). It is important to note, that the low-energy physics of the ABCA configuration is entirely dominated by the A sites of the topmost and the B sites of the bottomost layers. Since correlations have a negligible  effect on the ABAB regions, we will concentrate our discussion of correlations on the ABCA domains in the following. 
	
	\begin{figure}[t]
		\centering
		\includegraphics[width=\linewidth]{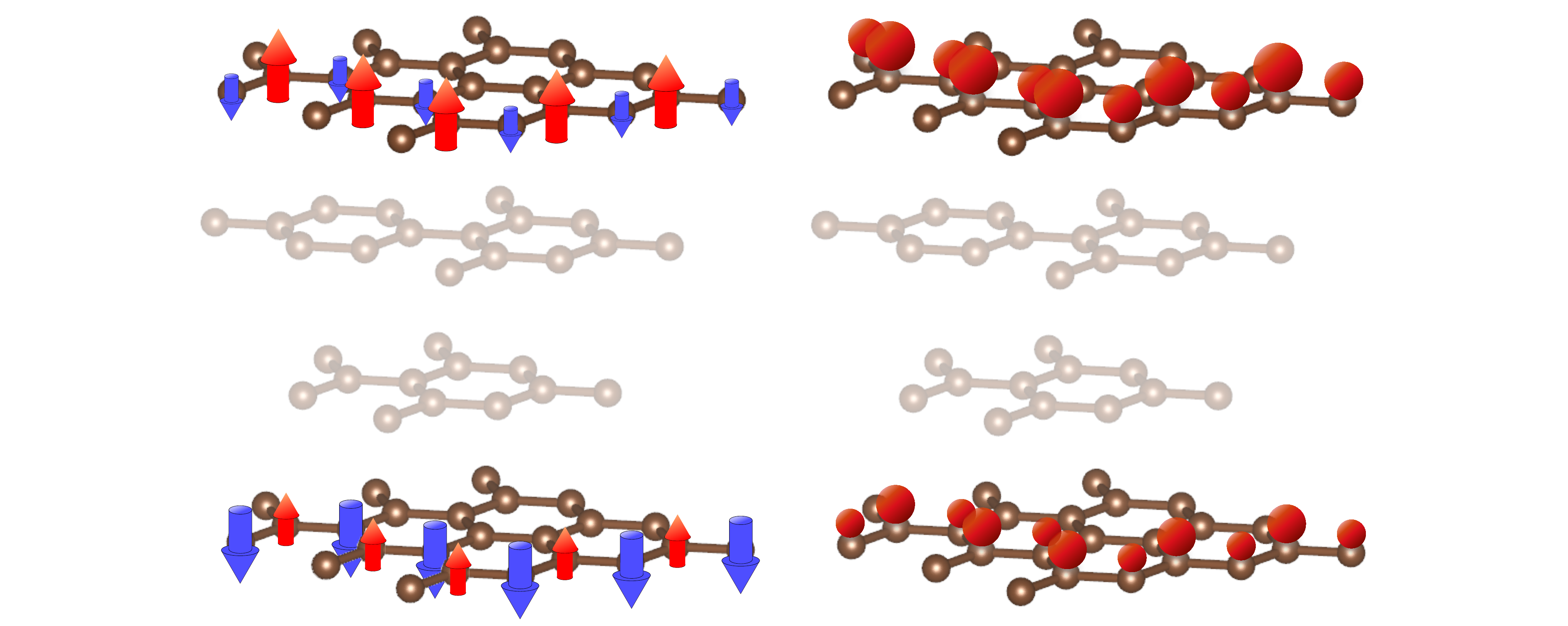}
		\caption{Visual representation of the order in ABCA stacked bi-bi-layer. The mostly irrelevant middle layers are shaded. Ferrimagnetic spin order (left) and charge transfer insulating order (right) are shown by arrows pointing up or down as well as big or small spheres, respectively (shown on select sites only).    }
		\label{fig:Order}
	\end{figure}
	
	We do not have an ab-initio or experimental characterization of the precise form and magnitude of Coulomb matrix element. We thus concentrate on the most short-ranged interactions within a layer as well as across topmost to bottommost layer on A and B sites, respectively, as those are the only ones that contribute to the low-energy physics. We therefore concentrate on adding an additional contribution $U' \sum_i\sum_{\sigma,\sigma'} \left(n_{i,1A,\sigma}-\frac{1}{2}\right)\left(n_{i,4B,\sigma'}-\frac{1}{2}\right)$ to $H_U$ to explore the general physics keeping $U$ and $U'$ as parameters to be swept. When turning on $U'$, at values $U'\approx U$, the ferrimagentic state discussed above (at $U'=0$) gives way to a charge transfer (excitonic) insulating state which spontaneously breaks the inversion symmetry between bottom and top layer, spontaneously transferring charge across the $1A$ and $4B$ sites. In contrast to the Ferrimagnetic state the charge transfer state is compatible with an external magnetic field (which sets an explicit symmetry breaking of the same kind as the spontaneous one of the charge transfer order). A visual representation of these orders in ABCA stacked bi-bi-layer is depicted in Fig.~\ref{fig:Order}.
	
	Due to the spontaneous charge transferred, the local density of states at the 1A sites becomes highly asymmetric with respect to $\omega$ in contrast to the local density of states in a ferrimagnetically ordered state (see above). We give a full sweep through the $U$-$U'$ plane keeping $E_f=0$ in Fig.~\ref{fig:Sweep}, where we show the gap size (splitting of van-Hove singularities)  on the left and the asymmetry (indicating the charge transferred) on the right. Furthermore, we indicate the line of $U$-$U'$ values where the gap size is about the experimentally reported $10$meV.
	
	\begin{figure}[t]
		\centering
		\includegraphics[width=.32\columnwidth]{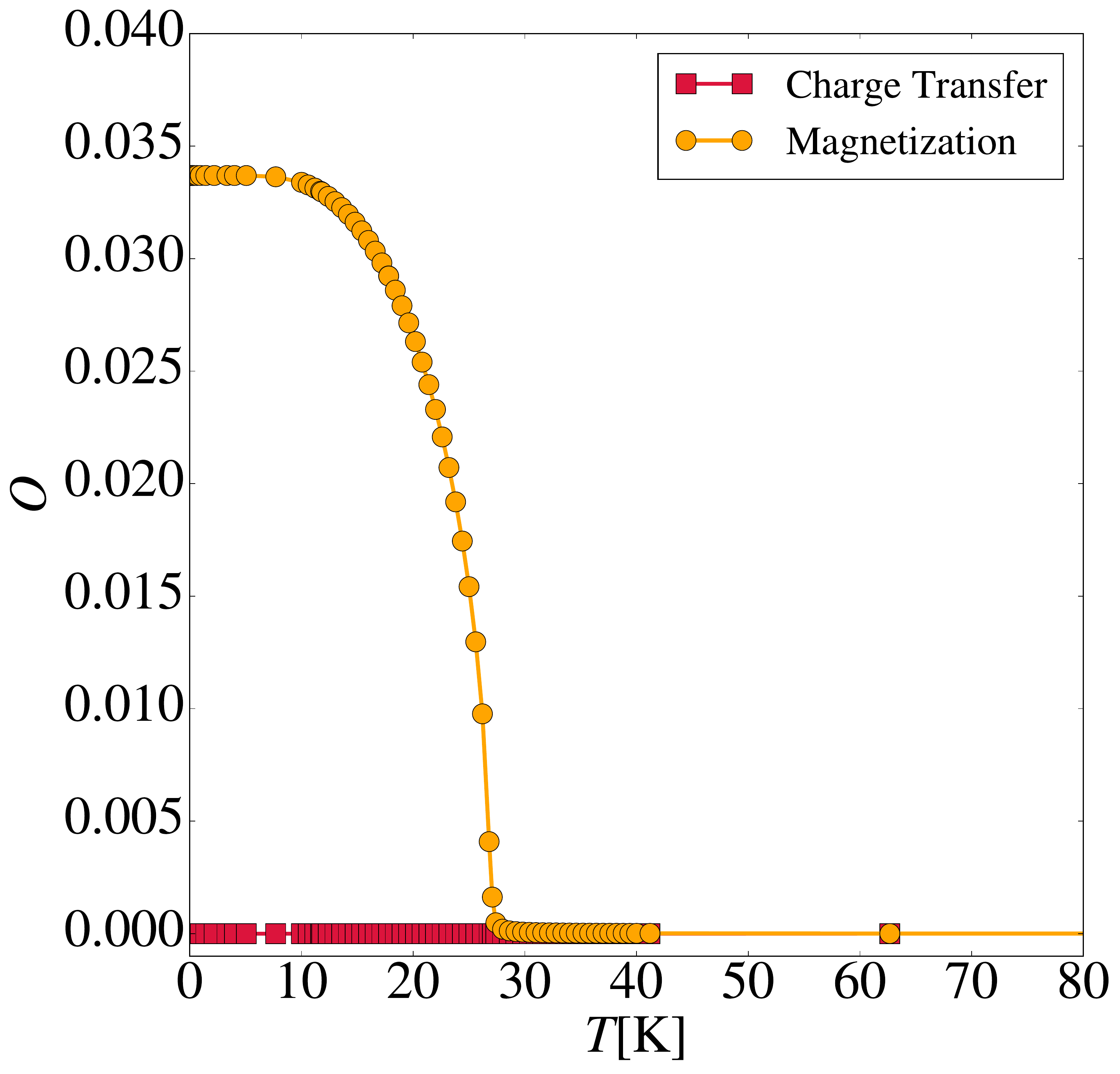}
		\includegraphics[width=.32\columnwidth]{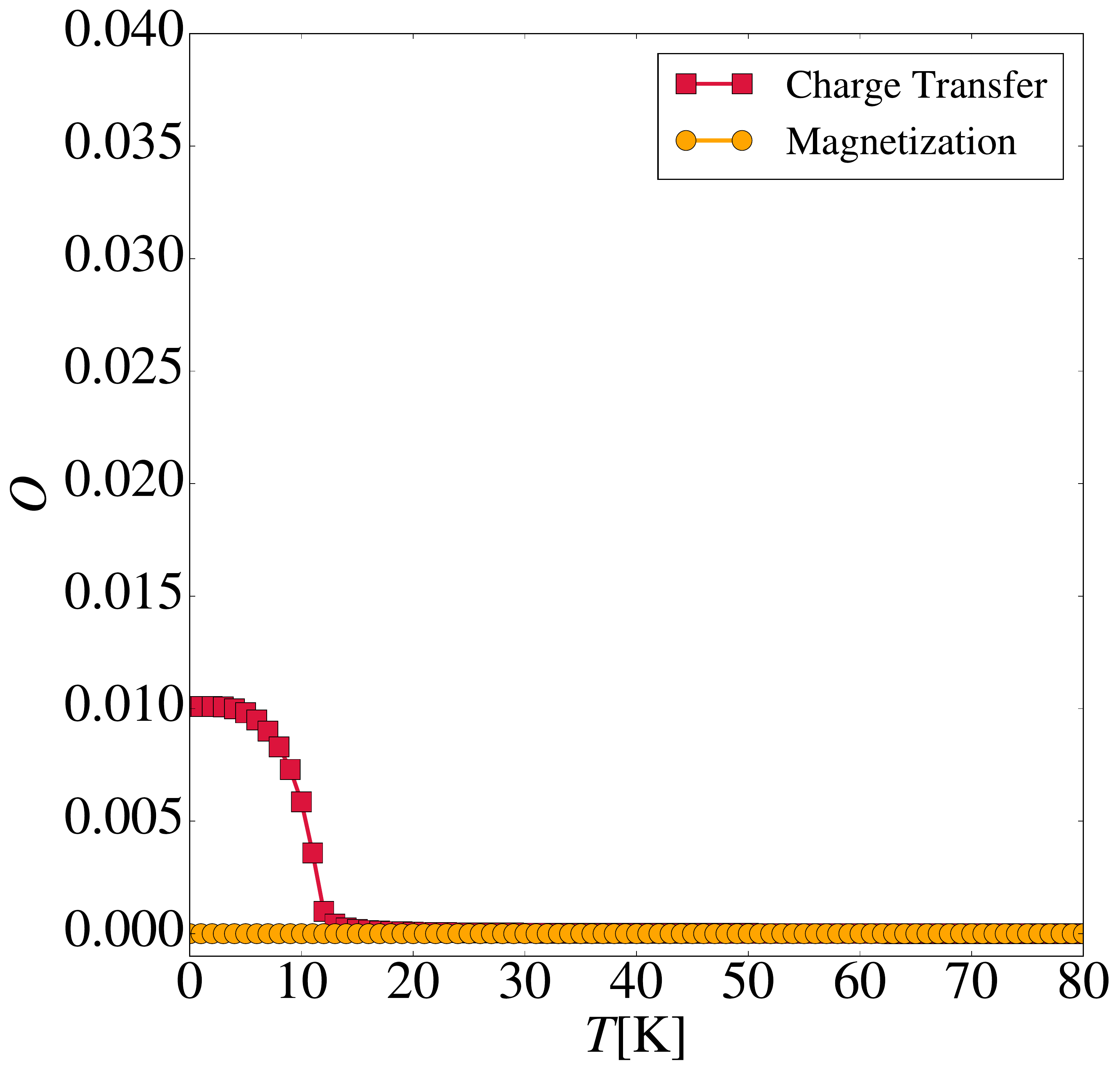}
		\caption{ Temperature dependence of the order parameters for the magnetic and charge order for $U=0.275$eV and $U'=0$ (left) as well as $U=0.1$eV and $U'=0.2$eV (left) at $E_f=0$. The mean field critical temperature is approximately $T_C\approx28K$ and  $T_C\approx12K$ for the two cases.       }
		\label{fig:DIAT}
	\end{figure}
	
		\begin{figure}[t]
		\centering
		\includegraphics[width=.5\columnwidth]{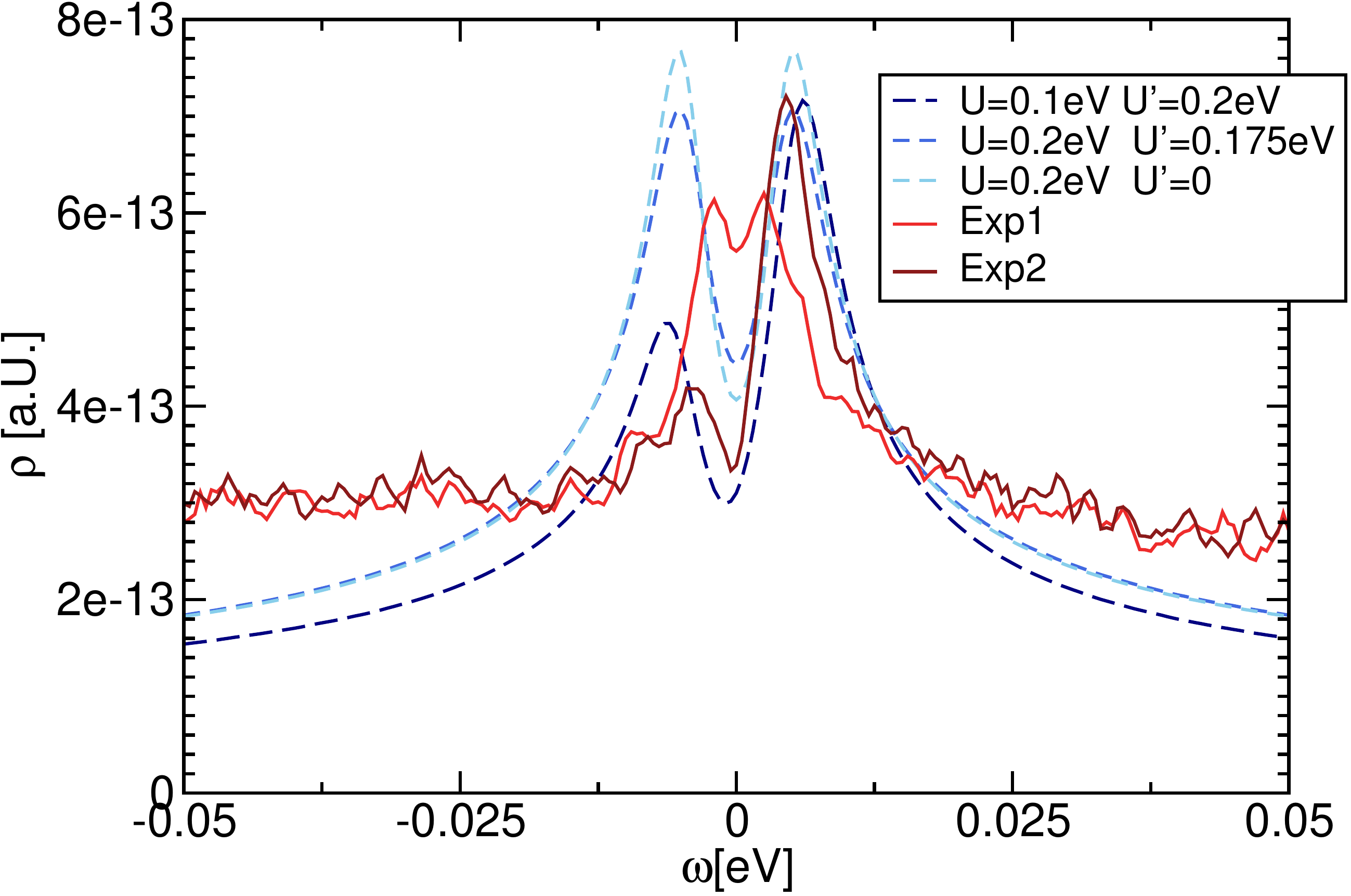}
		\caption{Comparison of theory to experiment. Experimentally, we concentrate on the most symmetric scenarios for the density of states (labeled Exp2) and the case which is determined to be closest to half-filling and zero applied displacement field (labeled Exp1). The splitting size between the van-Hove singularities is roughly fitted by varying $U$ and $U'$. The experimental data has a strong tendency to display asymmetric peaks. This indicates that the experimental sample might be (at the verge of) forming spontaneously a charge transfer state even at small $E_f$.       }
		\label{fig:comp}
	\end{figure}

	Next, we analyze the electric field dependence. In Fig.~\ref{fig:DOS_E} we show the density of states on the A sites of the topmost layer for electric fields from $E_f=-0.2eV$ to $E_f=0.2eV$ sampled in increments of $0.01eV$. We choose three combinations of $(U,U')=(0.2eV,0.0)$ (left), $(U,U')=(0.275eV,0.175eV)$ (center) and $(U,U')=(0.1eV,0.2eV)$ (right) on the dashed line shown in the left panel of Fig.~\ref{fig:Sweep} denoting a splitting of van-Hove singularities of roughly $10$meV. Electric field, just as the charge transfer state stabilized by $U'$ tends to separate charges across the bottommost and topmost layer. Therefore at non-zero $U'$ the electric field induced gap in the density of states (and the charge transfer) is stronger.
	
	Next we define the two order parameters for (ferri)magnetic 
	\begin{equation}
	O^{\rm Mag}=\frac{N_{1A,\uparrow}-N_{1A,\downarrow}-N_{4B,\uparrow}+N_{4B\downarrow}}{N_{1A,\uparrow}+N_{1A,\downarrow}+N_{4B,\uparrow}+N_{4B,\downarrow}}
	\end{equation}
	as well as charge transfer
	\begin{equation}
	O^{\rm CT}=\frac{N_{1A,\uparrow}-N_{4B,\uparrow}+N_{1A,\downarrow}-N_{4B,\downarrow}}{N_{1A,\uparrow}+N_{1A,\downarrow}+N_{4B,\uparrow}+N_{4B,\downarrow}}\end{equation}
	order, where $N_{a,\sigma}=\sum_i n_{i,a,\sigma}$. Results for these order parameters are shown in Fig.~\ref{fig:SUS_E} for $(U,U')=(0.2eV,0.0)$ (left), $(U,U')=(0.275eV,0.175eV)$ (center) and $(U,U')=(0.1eV,0.2eV)$ (right). Charge transfer competes with the magnetic order and is stabilized by finite electric field as one would expect. At large $U'>U$ charge transfer can be spontaneous breaking the underlying inversion symmetry (see right panel). 
	
	The temperature dependence of the order parameter shows the typical mean-field behavior illustrated in Fig.~\ref{fig:DIAT} for $U=0.275$eV and $U'=0$ (left panel) as well as for $U=0.1$eV and $U'=0.2$eV (right panel) at $E_f=0$. For $U=0.275$eV and $U'=0$ magnetic ordering sets in below a critical temperature of $T_C=28K$ while there is no charge order for this choice of parameters. On the other hand for $U=0.1$eV and $U'=0.2$eV charge ordering emerges at $T_C\approx 12K$ without magnetic ordering.

	While experimentally we see that the charge transferred order is more likely due to the asymmetry of the peaks, we do not think that this is definitive evidence for charge transferred order as opposed to Ferrimagnetism -- this is left to investigation by future works. Here we take a simple vantage point and compare the most symmetric density of states obtained in the experiment as well as the density of states of the point of near-zero applied displacement field with the $E_f=0$. Results of the mean field calculation at different combinations of $U$ and $U'$ which roughly reproduce the experimental gap of $10$meV are summarized in Fig.~\ref{fig:comp}. Our analysis suggests that the experimental system is likely showing spontaneous charge transfer order.
	
	\section*{ABAB Field Dependent STS}

\begin{figure*}[t]
	\includegraphics[width=\linewidth]
	{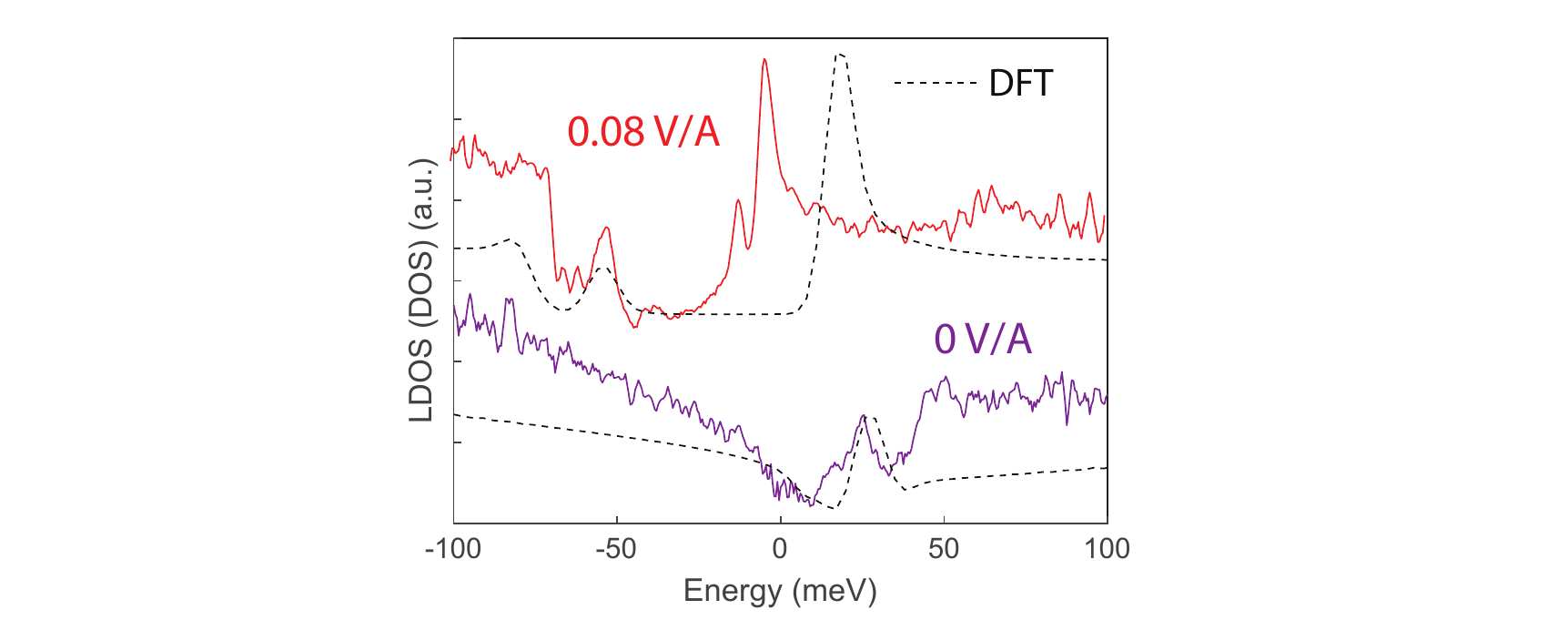}
	\caption{
	STS LDOS and DFT top layer DOS at 0 and 0.8 V/nm displacement fields in ABAB graphene.} 
	\label{fig:S4}
\end{figure*}

	Figure \ref{fig:S4} shows STS LDOS curves on an ABAB domain at 0 V/nm and 0.8 V/nm along with DFT calculated top layer DOS. The experimental and theoretical curves match nicely confirming the ABAB band structure and that a gap opens on the top layer under an applied displacement field.
	
	\bibliographystyle{unsrtnat}
	\bibliography{tdbgbibtexsupp}